\documentclass{iopart}
\usepackage{euscript,iopams,amssymb,amsfonts,graphicx,bm}
\usepackage{pgfplots,setstack}

\usepackage{float}
\usepackage{epsfig}
\usepackage{esint}
 \usepackage{tikz-cd}

\usepackage{bm,braket}

\eqnobysec

\renewcommand{\theequation}{\arabic{section}.\arabic{equation}}

\newcommand{\calU}{{\mathcal U}}

\newcommand{\calP}{{\mathfrak P}}

\newcommand{\calS}{{\mathfrak S}}
\newcommand{\calJ}{{\mathcal J}}
\newcommand{\calPP}{\widetilde{\mathfrak P}}
\newcommand{\calQQ}{\widetilde{\mathfrak Q}}

\newcommand{\R}{{\mathbb R}}

\renewcommand{\P}{\mathbb{P}}

\newcommand{\x}{\mathbf{x}}

\renewcommand{\e}{{\mathrm e}}
\newcommand{\E}{{\mathbb E}}
\newcommand{\pcb}[1]{\textcolor{black}{#1}}

\newcommand{\calT}{{\mathcal T}}
\renewcommand{\P}{\mathbb P}
\newcommand{\p}{\widetilde{p}}

\begin{document}

 \title[Encounter-based model of a run-and-tumble particle with resetting]{Encounter-based model of a run-and-tumble particle with stochastic resetting}

\author{Paul C. Bressloff}
\address{Department of Mathematics, Imperial College London, 
London SW7 2AZ, UK}

\begin{abstract} 
In this paper we analyze the effects of stochastic resetting on an encounter-based model of an unbiased run-and-tumble particle (RTP) confined to the half-line $[0,\infty)$ with a partially absorbing wall at $x=0$. The RTP tumbles at a constant rate $\alpha$ between the velocity states $\pm v$ with $v>0$. Absorption occurs when the number of collisions with the wall (discrete local time) exceeds a randomly generated threshold $\widehat{\ell}$ with probability distribution $\Psi(\ell)$. The extended RTP model has three state variables, namely, particle position $X(t)\in [0,\infty)$, the velocity direction $\sigma(t)\in\{-1, 1\}$, and the discrete local time $L(t)\in {\mathbb N}$. We initially assume that only $X(t)$ and $\sigma(t)$ reset at a Poisson rate $r$, whereas $L(t)$ is not changed. This implies that resetting is not governed by a renewal process. We use the stochastic calculus of jump processes to derive an evolution equation for the joint probability distribution of the triplet $(X(t),\sigma(t),L(t))$. This is then used to calculate the mean first passage time (MFPT) by performing a discrete Laplace transform of the evolution equation with respect to the local time. We thus find that the MFPT's only dependence on the distribution $\Psi$ is via the mean local time threshold. We also identify parameter regimes in which the MFPT is a unimodal function of both the resetting and tumbling rates. Finally, we consider conditions under which resetting is given by a renewal process and show how the MFPT in the presence of local time resetting depends on the full statistics of $\Psi$.

\end{abstract}

\maketitle
\section{Introduction}

Encounter-based models of diffusion-mediated surface reactions assume that the probability of absorption depends upon the amount of particle-surface contact time \cite{Grebenkov20,Grebenkov20a,Grebenkov22}. The latter is determined by a Brownian functional known as the boundary local time $\ell(t)$ \cite{Ito65,McKean75}. Encounter-based models can also be applied to diffusion in domains with partially absorbing interior traps \cite{Bressloff22,Bressloff22a}. In this case, a particle freely enters and exits a trap, but can only be absorbed within the trapping region, so that the probability of absorption depends on the particle-trap encounter time. The latter is equivalent to the Brownian occupation time. There are now a wide range of search-and-capture processes that have been studied using the encounter-based approach. These include multiple targets \cite{Bressloff22e,Grebenkov22a},
subdiffusive processes \cite{Bressloff23b,Bressloff23c}, diffusion processes with stochastic resetting \cite{Bressloff22b,Bressloff22c,Benk22}, and non-diffusive processes such as active run-and-tumble particles (RTPs) \cite{Bressloff22d,Bressloff23}. In the last example the particle-surface contact time is the amount of time the particle spends ``stuck'' to the boundary. In the case of a non-sticky boundary, this reduces to the number of collisions with the boundary (discrete local time). Finally, diffusion through a semi-permeable membrane can also be generalized using an encounter-based model \cite{Bressloff22f,Bressloff23d} by formulating sample paths in terms of so-called snapping out Brownian motion \cite{Lejay16}.

There are two main features of encounter-based models that underly their wide applicability. First, absorption at a surface or within an interior trap is separated from the bulk dynamics such that the probability of absorption only depends on the particle-target contact time. Second, an absorption event is identified as the first time that the contact time crosses a randomly generated threshold. Different models of absorption (Markovian and non-Markovian) then correspond to different choices of the random threshold probability distribution $\Psi$. Markovian absorption assumes that the probability of absorption over an infinitesimal contact time increment is independent of the accumulated contact time (memoryless absorption). This corresponds to an exponential density $\Psi$ in the case of a Brownian particle, whereas $\Psi$ is a geometric distribution in the case of a non-sticky RTP \cite{Bressloff22d}. Non-Markovian absorption occurs when the probability of absorption depends on some internal state of the absorbing substrate or particle, and activation/deactivation of this state proceeds progressively by repeated particle-target encounters \cite{Bartholomew01,Filoche08}. A general method for analyzing an encounter-based model is to determine the joint probability density or generalized propagator for a particle's state (position, velocity etc.) and the target contact time. This can be achieved by solving a Robin-like boundary value problem (BVP) for a constant rate of absorption $\kappa_0$. The constant $\kappa_0$ is then reinterpreted as a Laplace variable $z$ conjugate to the contact time, and the inverse Laplace transform of the classical solution with respect to $z$ yields the propagator. Performing a $\Psi$-dependent weighted integration of the propagator with respect to the contact time then determines the marginal probability density for particle position etc.

In this paper we consider an encounter-based model of an RTP on the half-line with stochastic resetting and a partially absorbing (non-sticky) wall. (Previous related work on RTPs has focused on first passage time problems (FTPs) in the case of stochastic resetting and totally absorbing walls \cite{Evans18,Bressloff20,Santra20a} or partially absorbing walls without resetting \cite{Angelani15,Angelani17,Angelani23,Malakar18,Demaerel18,Bressloff22d,Bressloff23}.) The extended RTP model has three state variables, namely, particle position $X(t)\in [0,\infty)$, the velocity direction $\sigma(t)=\in\{-1, 1\}$, and the discrete local time (number of wall collisions) $L(t)$. We can then distinguish different resetting protocols based on which subset of variables are reset. In this paper we assume that both $X(t)$ and $\sigma(t)$ reset but allow for the possibility that $L(t)$ does not. As we have previously highlighted for diffusion processes \cite{Bressloff22b,Bressloff22c}, the latter makes sense when there are contact-dependent changes in the activation/deactivation state of the target rather than the particle prior to absorption. In contrast to most studies of stochastic resetting \cite{Evans20}, the resulting resetting protocol is not governed by a renewal process. One way to recover the renewal property is to assume that the memory trace for absorption is encoded by an internal state of the particle that varies monotonically with $L(t)$. Resetting the internal state then effectively resets $L(t)$.

The structure of the paper is as follows. In section 2 we briefly recap the analysis of the FTP problem for an RTP with resetting on the half-line in the presence of a totally absorbing wall \cite{Evans18}. The encounter-based model of an RTP for a partially absorbing wall is then presented in section 3, first in the absence of resetting and then with resetting. The corresponding Chapman-Kolmogorov (CK) equations for the discrete local time propagator are derived from first principles using the stochastic calculus of run-and-tumble motion. The detailed calculation of the MFPT for absorption is carried out in section 4 and appendix A, under the assumption that the local time does not reset. We find that the MFPT's only dependence on the local time threshold distribution $\Psi$ is via the mean threshold. We also identify parameter regimes in which the MFPT is a unimodal function of both the resetting and tumbling rates. Finally, in section 5, we consider conditions under which resetting is given by a renewal process and show how the MFPT in the presence of local time resetting depends on the full statistics of $\Psi$. 

\section{RTP on the half-line with a totally absorbing wall}

As a point of comparison, we begin by briefly reviewing a previous analysis of the first passage time (FTP) problem for an RTP confined to the half-line $x\in [0,\infty)$ with a totally absorbing wall at $x=0$ \cite{Evans18}. The RTP switches between two constant velocity states $v_+=v$ and $v_-=-v$, $v>0$, according to a Poisson process with rate $\alpha$. Prior to hitting the wall, the position $X(t)$ of the particle at time $t$ evolves according to the piecewise deterministic equation
\begin{equation}
\label{PDMP}
\frac{dX}{dt}=v\sigma(t),
 \end{equation}
where $\sigma(t)=\pm 1$ is a dichotomous noise process that switches sign at the rate $\alpha$. Let $p_{k}(x,t)$ be the probability density of the RTP at position $x\in \R^+$ and velocity state $\sigma(t)=k$. The associated differential Chapman-Kolomogorov (CK) equation is \cite{Evans18}
\numparts
\begin{eqnarray}
\label{DLa}
\frac{\partial p_{1}}{\partial t}&=-v \frac{\partial p_{1}}{\partial x}-\alpha p_{1}+\alpha p_{-1},\\
\frac{\partial p_{-1}}{\partial t}&=v \frac{\partial p_{-1}}{\partial x}-\alpha p_{-1}+\alpha p_{1}.
\label{DLb}
\end{eqnarray}
This is supplemented by the absorbing boundary condition $p_1(0,t)=0.$
\endnumparts
The initial conditions are taken to be $x(0)=x_0>0$ and $\sigma(0)=\pm 1$ with probability $\rho_{\pm 1}=1/2$.

Now suppose that the position $X(t)$ is reset to its initial location $x_0$ at random times generated from a Poisson distribution with constant rate $r\geq 0$. Following Ref. \cite{Evans18}, we also assume that the discrete state $\sigma(t)$ is reset to $\sigma_0=\pm 1$ with probability $\rho_{\pm 1}=1/2$ (velocity randomization). 
The resulting probability density with resetting satisfies the modified CK equation
\numparts
\begin{eqnarray}
\label{CKHa}
\frac{\partial p_{1}}{\partial t}&=-v \frac{\partial p_{1}}{\partial x}-(\alpha +r) p_{1}+\alpha  p_{-1}
+\frac{rS(t)}{2}\delta(x-x_0) ,\\
\label{CKHb}
\frac{\partial p_{-1}}{\partial t}&=v \frac{\partial p_{-1}}{\partial x}-(\alpha +r) p_{-1}+\alpha  p_{1}
+\frac{rS(t)}{2}\delta(x-x_0), \\
 {p}_1(0,t)&=0,
 \label{CKHc}
\end{eqnarray}
\endnumparts
where $S(t)$ is the survival probability
 \begin{equation}
 S(t)=\int_0^{\infty}p(x,t)dx
 \end{equation}
and
\begin{equation}
p(x,t)=p_1(x,t)+p_{-1}(x,t),\quad J(x,t)=v[p_1(x,t)-p_{-1}(x,t)].
\end{equation}
 If $\calT=\inf\{t>0,\ X(t)=0\}$ denotes the first passage time (FPT) for absorption, then ${\rm Prob}[T\leq t] =1-S(t)$. 
  The FPT density is thus given by
\begin{equation}
 f(t)=-\frac{\partial S(t)}{\partial t}= -J(0,t)
\end{equation}
and the MFPT is 
\begin{eqnarray}
  \tau &:=  \int_0^{\infty}t f (t)dt 
=\lim_{s\rightarrow 0}\widetilde{S} (s),
\end{eqnarray}
where $\widetilde{S} (s)=\int_0^{\infty}\e^{-st}S(t)dt$.

The survival probability satisfies the renewal equation \cite{Evans18}
\begin{equation}
\label{ren}
S(t)=\e^{-rt} {S}_0(t)+r\int_0^{\infty} \e^{-r\tau} {S}_0(\tau) S(t-\tau)d\tau,
\end{equation}
where ${S}_0(t)=\left [ \int_0^{\infty}p(x,t)dx \right ]_{r=0}$ is the survival probability without resetting. The first term on the right-hand side represents the contribution from survival trajectories that have no resettings, while the second term sums over the contributions from survival trajectories that last reset at time $t-\tau$. Laplace transforming the last renewal equation and rearranging implies that 
\begin{equation}
\label{ren2}
\widetilde{S}(s)=\frac{\widetilde{S}_0(s+r)}{1-r\widetilde{S}_0(s+r)}=-\frac{1}{r} +\frac{1}{r}\frac{1}{1-r\widetilde{S}_0(s+r)}.
\end{equation}
In Ref. \cite{Evans18} the survival probability $\widetilde{S}_0(s)$ is calculated by solving the Laplace transformed backwards CK equation. One finds that
\begin{equation}
\widetilde{S}_0(s)=\frac{1}{s}+\frac{1}{2\alpha s}(v\Lambda_0(s)-(s+2\alpha)]\e^{-\Lambda_0(s)x_0},
\end{equation}
with
\begin{equation}
\Lambda_0(s)\equiv \frac{1}{v}\sqrt{s(2\alpha+s )}.
\label{Lam}
\end{equation}
Substituting into equation (\ref{ren2}) then yields the result
\begin{equation}
\fl \widetilde{S}(s)=-\frac{1}{r}+\frac{1}{r}\left [\frac{2\alpha (s+r)\e^{\Lambda_0(s+r)x_0}}{2\alpha s\e^{\Lambda_0(s+r)x_0}-r[v\Lambda_0(s+r)-(r+s+2\alpha)]}\right ].
\end{equation}
Finally, taking the limit $s\rightarrow 0$ leads to the following expression for the MFPT \cite{Evans18}:
\begin{equation}
\tau=-\frac{1}{r}+\frac{2\alpha}{r}\left [\frac{ \e^{\Lambda_0(r)x_0}}{r+2\alpha-\sqrt{r(2\alpha+r )}}\right ].
\label{sym0}
\end{equation}
\setcounter{equation}{0}

\section{RTP on the half-line with a partially absorbing wall}

Now suppose that there is a partially absorbing wall at $x=0$. In previous work we used an encounter-based probabilistic framework to analyze the FPT problem for an RTP in a finite interval with partially absorbing walls and no resetting \cite{Bressloff22}. The basic idea of encounter-based methods is that the probability of absorption depends on the amount of contact time between the particle and wall \cite{Grebenkov20,Grebenkov22,Bressloff22,Bressloff22a}. In this section we extend the probabilistic framework to incorporate stochastic resetting on the half-line. (Resetting generates a finite MFPT on the half-line, so we can consider this simpler configuration.) In order to proceed, we look more closely at the SDE underlying run-and-tumble motion and use the stochastic calculus of jump processes.

Let $\calT_n$, $n \geq 1$, denote the $n$th time that the RTP reverses direction and introduce the inter-reversal times $\tau_n=\calT_{n}-\calT_{n-1}$ with
\begin{equation}
\P[\tau_n\in [s,s+ds]]=\alpha \e^{-\alpha s}ds.
\end{equation} 
 The number of direction reversals occurring in the time interval $[0,t]$ is given by the Poisson process
 $N(t)$ with
 \begin{eqnarray}
N(t)=n,\quad \calT_{n}\leq t <\calT_{n+1}.
\end{eqnarray}
Note that $N(t)$ is defined to be right-continuous. That is, $N (\calT_n^-)=n-1$ whereas $N(\calT_n)=n$.
The probability distribution of the Poisson process is
\begin{eqnarray}
\P[N(t)=n]=  \frac{(\alpha t)^{n}\e^{-\alpha t}}{n!},
\label{Pois1}
\end{eqnarray}
with $\E[N(t)]=\alpha t=\mbox{Var}[N(t)]$. Away from the boundary at $x=0$, the position $X(t)$ of the particle at time $t$ evolves according to the SDE 
\begin{equation}
\label{PDMP0}
dX(t)=\sigma(t)vdt,\quad d\sigma(t)=-2\sigma(t^-)dN(t),
 \end{equation}
where $\sigma(t)=\pm 1$ is a dichotomous noise process with
\begin{equation}
 dN(t)=h(t)dt,\quad h(t)=\sum_{n=1}^{\infty} \delta(t-\calT_n).
\end{equation}
We will assume throughout that the initial state is $X(0)=x_0>0$ and $\sigma(0)=\sigma_0$ (in contrast to velocity randomization). For the moment we do not include stochastic resetting.

\subsection{Partial absorption and the discrete local time}First suppose that the boundary at $x=0$ is totally reflecting. This means that each time $t$ the particle hits the boundary $X(t)=0$ in the state $\sigma(t)=-1$, there is an instantaneous reversal in direction. This can be implemented by introducing a second right-continuous counting process $L(t)$, in which $L(t)$ is the number of collisions with the wall over the time interval $[0,t]$. (In general $L(t)$ is non-Poissonian.) We can then set 
\begin{equation}
\label{loc}
L(t) =v \int_0^t\delta(X(\tau))d\tau   =\int_0^t \sum_{n\geq 1}\delta(\tau-S_n)d\tau,
\end{equation}
where $S_n$ is the time of the $n$th collision with the wall. 
Each trajectory of the RTP will typically yield a different value of $L(t)$, which means that $L(t)$ is itself a stochastic process. Note that $L(t)$ plays the discrete analog of the local time for Brownian motion \cite{Singh21,Bressloff22}.
The reflecting boundary can be incorporated into equation (\ref{PDMP0}) by taking
\numparts
\begin{equation}
\label{PDMPa}
dX(t)=\sigma(t)vdt,\quad d\sigma(t)=-2\sigma(t^-)[dN(t)+ dL(t)],
 \end{equation}
 where
 \begin{equation}
 \label{PDMPb}
dL(t)= \eta(t) dt,\quad \eta(t)=v\delta(X(t))=\sum_{n\geq 1}\delta(t-S_n). 
 \end{equation}
 \endnumparts
 That is, there is either a spontaneous reversal of direction at a Poisson rate $\alpha$ anywhere in the bulk domain or a forced reversal due to reflection at the wall.

Given the definition of the discrete local time for an RTP, we can impose a partially absorbing boundary condition by introducing the stopping time condition \cite{Bressloff22}
\begin{equation}
\label{TA}
{\mathcal T}=\inf\{t>0:L(t) >\widehat{\ell}\},
\end{equation}
where $\widehat{\ell}$ is a random variable with probability distribution 
 \begin{equation}
 \psi(\ell)=\P[\widehat{\ell}=\ell],\quad \sum_{\ell=0}^{\infty}\psi(\ell)=1.
 \end{equation}
We also have the relations
 \begin{equation}
\Psi(\ell) \equiv \P[\widehat{\ell}>\ell]=\sum_{\ell'=\ell+1}^{\infty} \psi(\ell')=1-\sum_{\ell'=0}^{\ell}\psi(\ell').
\end{equation}
and
\begin{equation}
\psi(\ell)=\Psi(\ell-1)-\Psi(\ell).
\label{psi}
\end{equation}
Note that ${\mathcal T}$ is a random variable that specifies the first absorption time, which is identified with the event that $L(t)$ first exceeds a randomly generated threshold $\widehat{\ell}$. 

For a given threshold function $\Psi(\ell)$, the joint probability density for particle position and the velocity state can be expressed as
\begin{equation}
 p_{k}^{\Psi}(x,t)dx=\P[X(t) \in (x,x+dx), \ \sigma(t)=k,\ t < {\mathcal T}].
\end{equation}
Given that $L(t)$ is a nondecreasing process, the condition $t < {\mathcal T}$ is equivalent to the condition $L(t)\leq  \widehat{\ell}$. This implies that
\begin{eqnarray*}
\fl p_{k}^{\Psi}(x,t)dx&=\P[X(t) \in (x,x+dx),\ \sigma(t)=k, L(t) \leq \widehat{\ell}]\\
\fl &=\sum_{\ell=0}^{\infty} \psi(\ell)\P[X(t) \in (x,x+dx),\ \sigma(t)=k , \ L(t) =\ell]\\
\fl &\equiv\sum_{\ell=0}^{\infty} \psi(\ell)\sum_{m=0}^n  P_{k,\ell}(x,t)\, dx,
\end{eqnarray*}
where $  P_{k,\ell}(x,t)$ is the analog of the local time propagator for diffusion \cite{Grebenkov20,Bressloff22}.
Using the identity
\begin{equation}
\sum_{n=0}^{\infty}f(n)\sum_{m=0}^n   g(m)=\sum_{m=0}^{\infty}g(m)\sum_{n=m}^{\infty}  f(n) 
\label{fg}
\end{equation}
for arbitrary functions $f,g$, it follows that
\begin{equation}
\label{bob}
p_{k}^{\Psi}(x,t)=\sum_{\ell=0}^{\infty}\Psi(\ell-1)P_{k,\ell}(x,t),
\end{equation}
with $\Psi(-1)=1$.

\subsection{CK equation for the discrete local time propagator}
In a previous paper we showed that the local time propagator $P_{k,\ell}(x,t)$ evolves according to a CK equation, which on the half-line takes the explicit form \cite{Bressloff22}
\numparts
\begin{eqnarray}
\label{CKpropa}
  &\frac{\partial P_{k,\ell}(x,t)}{\partial t}=-vk\frac{\partial P_{k,\ell}(x,t)}{\partial x}+\alpha [P_{-k,\ell}(x,t)-P_{k,\ell}(x,t)],
\\
 &P_{1,\ell+1}(0,t)=P_{-1,\ell}(0,t), \quad \ell \geq 0.
 \label{CKpropb}
\end{eqnarray}
\endnumparts
for $\ell\geq 0, \ k=\pm 1,\ x \in \R$.
However, we used a heuristic argument to determine the local time boundary condition (\ref{CKpropb}). Here we derive the propagator CK equation from first principles using the stochastic calculus of jump processes. 

Consider the SDE given by equations (\ref{PDMPa}) and (\ref{PDMPb}).
 Let $f$ be an arbitrary smooth bounded function on $\R^+\times\R^+\times \{-1,1\}$.  Applying It\^o's formula for  jump processes we have
\begin{eqnarray}
 \fl  df(X(t),L(t),\sigma(t)) 
 &= v\sigma(s) \partial_xf(X(t),L(t),\sigma(t)) dt \\
  \label{Ito}
\fl  &\quad +\bigg [f(X(t),L(t),-\sigma(t^-))-f(X(t),L(t),\sigma(t^-))\bigg ]dN(t)\nonumber \\
\fl &\quad+\bigg [f(X(t),L(t^-)+1,-\sigma(t^-))-f(X(t),L(t^-),\sigma(t^-))\bigg ] dL(t).\nonumber \end{eqnarray}
Introducing the empirical measure
\begin{equation}
  \rho_{k,\ell}(x,t)=\delta(x-X(t))\delta_{L(t),\ell} \delta_{k,\sigma(t)}
\end{equation}
for $X(t)\in [0,\infty)$ and $k=\pm 1$, we have the identity
\begin{equation}
 \sum_{\ell\geq 0} \sum_{k=\pm 1} \int_0^{\infty}  \rho_{k,\ell}(x,t)f(x,\ell,k)dx=F(t)\equiv f(X(t),L(t),\sigma(t)).
\end{equation}
Taking differentials of both sides with respect to $t$ yields
\begin{eqnarray}
\label{ffa}
 \left \{ \sum_{\ell \geq 0} \sum_{k=\pm 1} \int_{0}^{\infty}  f(x,\ell,k)\frac{\partial \rho_{k,\ell}(x,t)}{\partial t}  dx\right \} = dF(t).
 \end{eqnarray}
  Using the generalised It\^o's lemma (\ref{Ito}), 
 \begin{eqnarray}
 \fl  &\left \{ \sum_{\ell \geq 0}\sum_{k=\pm 1}   \int_0^{\infty} f(x,\ell,k)\frac{\partial \rho_{k,\ell}(x,t)}{\partial t}  dx\right \}dt \nonumber\\
 \fl &\quad  =v  \sigma(t)\partial_x f(X(t),\sigma(t)) dt  +  [f(X(t),L(t^-),-\sigma(t^-))-f(X(t),L(t^-),\sigma(t^-)) ]dN(t)  \nonumber \\
 \fl&\qquad  +\bigg [f(X(t),L(t^-)+1,-\sigma(t^-))-f(X(t),L(t^-),\sigma(t^-))\bigg ] dL(t).
 \label{V1}
\end{eqnarray}
 From the definition of the empirical measure, 
 \begin{eqnarray}
   & \sum_{\ell \geq 0} \sum_{k=\pm 1}   \int_0^{\infty} f(x,\ell,k)\frac{\partial \rho_{k,\ell}(x,t)}{\partial t}  dx\nonumber \\
  &=v \sum_{\ell \geq 0}\sum_{k=\pm 1}  \int_{0} ^{\infty}k\rho_{k,\ell}(x,t) \partial_x f(x,\ell,k)   dx 
   \nonumber \\
    &\quad +h(t) \sum_{\ell \geq 0} \sum_{k=\pm 1}  \int_0^{\infty}   \rho_{k,\ell}(x,t^-)  [f(x,\ell,-k)-f(x,\ell,k)  ] dx \nonumber \\
  &\quad +\eta(t) \sum_{\ell \geq 0}  \int_0^{\infty}   \rho_{-1,\ell}(x,t^-)[f(x,\ell+1,1) -f(x,\ell,-1) dx.
  \label{V2}
\end{eqnarray}
Performing an integration by parts and resumming then gives
 \begin{eqnarray}
 \fl & \sum_{\ell\geq 0}\sum_{k=\pm 1}   \int_0^{\infty} f(x,\ell,k)\frac{\partial \rho_{k,\ell,}(x, t)}{\partial t}  dx\nonumber \\
 \fl &=-v\sum_{\ell\geq 0} \sum_{k=\pm 1}  \int_{0} ^{\infty}kf(x,\ell,k) \partial_x\rho_{k,\ell}(x, t)  dx-v \sum_{\ell\geq 0} \ \sum_{k=\pm 1}  kf(0,\ell,k) \rho_{k,\ell}(0, t) \nonumber  \\
 \fl  &\quad + h(t)\sum_{\ell\geq 0}\sum_{k=\pm 1}  \int_0^{\infty} f(x,\ell,k)  [\rho_{-k,\ell}(x, t^-)-\rho_{k,\ell}(x, t^-)\bigg ] dx .\nonumber  \\
 \fl &\quad +\eta(t)\sum_{\ell\geq 0} \ \sum_{k=\pm 1}  \int_0^{\infty}   f(x,\ell,k) \bigg [\rho_{-1,\ell-1}(x,t^-)\delta_{k,1}-\rho_{-1,\ell}(x,t^-)\delta_{k,-1}\bigg ] dx.
\end{eqnarray}
Using the identity $\eta(t)=v\delta(X(t))$ we can simplify the final term as
\[\sum_{\ell\geq 0} \ \sum_{k=\pm 1}   f(0,\ell,k) \bigg [\rho_{-1,\ell-1}(0,t^-)\delta_{k,1}-\rho_{-1,\ell,}(0,t^-)\delta_{k,-1}\bigg ] .\]
Since $f(x,\ell,k)$ and $f(0,\ell,k)$ are arbitrary, it follows that $\rho_k$ satisfies the SPDE (in the weak sense)
\begin{eqnarray}
  &\frac{\partial \rho_{k,\ell}}{\partial t}= -vk\frac{\partial}{\partial x} \rho_{k,\ell}(x, t)+h(t) [\rho_{-k,\ell}(x, t^-)-\rho_{k,\ell}(x, t^-)],
  \label{spuda}
  \end{eqnarray}
supplemented by the boundary condition 
\begin{equation}
\rho_{1,\ell+1}(0,t)=\rho_{-1,\ell}(0,t^-), \quad \ell \geq 0.
\label{spudb}
\end{equation}

The final step is to take expectations with respect to the Poisson process $N(t)$. Given some measurable function $F(x,\sigma)$, we have 
\begin{equation}
\E[F(X(t^-),\sigma(t^-))dN(t)]=\E[F(X(t^-),\sigma(t^-))]\E[dN(t)],
\end{equation}
since $F(X(t),\sigma(t))$ for all $t<\calT_n$ only depends on previous jump times. Moreover,
\begin{equation}
\fl \int_{\tau}^t\E[dN(s)] = \E[N(t)-N(\tau)]=\alpha (t-\tau)\mbox{ so that } \E[dN(t)]=\alpha dt.
\end{equation}
Taking expectations of equations (\ref{spuda})--(\ref{spudb}) and setting
\begin{equation}
P_{k,\ell}(x,a,t)=\E[\rho_{k,\ell}(x,t)],
\end{equation}
we obtain the propagator CK equations (\ref{CKpropa}) and (\ref{CKpropb}).

\subsection{Discrete Laplace transforms and the geometric distribution} Following Ref. \cite{Bressloff22}, suppose that $\Psi(\ell)$ is the geometric distribution:
\begin{equation}
\label{odd}
 \Psi(\ell-1)=z^{\ell}, \quad \psi(\ell)=(1-z)z^{\ell} , \quad \ell\geq 0
\end{equation}
Equation (\ref{bob}) implies that
\begin{eqnarray}
p_{k}(x,t)=\calP_k(x,z,t):=\sum_{\ell=0}^{\infty} z^{\ell} P_{k,\ell}(x,t|x_0), 
\end{eqnarray}
with $k=\pm 1,$ and $z\in [0,1]$. (For the geometric distribution we drop the superscript on $p_k^{\Psi}$.) That is, the probability density $p_k(x,t)$ is the discrete Laplace transform of the local time propagator. Moreover, 
the Laplace transformed propagator BVP is
\numparts
\begin{eqnarray}
\label{LTpropa}
\fl  \frac{\partial  \calP_k(x,z,t)}{\partial t}&=-vk\frac{\partial \calP_k(x,z,t)}{\partial x}+\alpha [ \calP_{-1}(x,z,t)-\calP_1(x,z,t)],
\label{LTpropb}\\
\fl &\calP_{1}(0,z,t)= z\calP_{-1}(0,z,t) .
\end{eqnarray}
\endnumparts
The boundary condition at $x=0$ is obtained as follows:
 \begin{eqnarray*}
 \fl0&=\sum_{\ell=0}^{\infty} z^{\ell}  [P_{1,\ell+1}(0,t)-P_{-1,\ell}(0,t)]\\
 \fl &=z^{-1}\sum_{\ell'=1}^{\infty} z^{\ell'}  P_{1,\ell'}(0,t)-\calP_{-1}(0,z,t)=\frac{1}{z}[ \calP_{1}(0,z,t)-z\calP_{-1}(0,z,t)],
\end{eqnarray*}
since $P_{1,0}(0,t)=0$. (There cannot be reversal of direction at the boundary until $L(t)>0$.)
Note that the boundary condition can be rewritten as
\begin{equation}
\calJ(0,z,t)\equiv v[\calP_1(0,z,t)-\calP_{-1}(0,z,t)]=-\kappa(z) \calP_10,z,t),
\label{calJ}
\end{equation}
where
\begin{equation}
\label{kaz}
\quad \kappa(z)= \frac{v[1-z]}{z}.
\end{equation}
 We conclude that
 the probability density $p_{k}(x,t)$ for an RTP on the half-line satisfying equations (\ref{DLa}) and (\ref{DLb}) and the Robin boundary condition
 \begin{equation}
 v[p_1(0,t)-p_{-1}(0,t)]=-\kappa_0 p_1(0,t)
 \end{equation}
 is equivalent to the discrete $z$-Laplace transform of the discrete local time propagator $P_{k,\ell}(x,t)$ with respect to the discrete local time $\ell$. The Laplace variable $z$ is related to $\kappa_0$ via equation (\ref{kaz}), that is, $z=z_0\equiv v/(v+\kappa_0)$.

Assuming that the inverse Laplace transform exists, we can incorporate a more general probability distribution $\Psi(\ell) $ according to
  \begin{equation}
  \label{Boo}
 \fl   p_{k }^{\Psi}(x,t)=\sum_{\ell=0}^{\infty} \Psi(\ell-1)P_{k,\ell}(x,t)=\sum_{\ell=0}^{\infty} \Psi(\ell-1){\mathbb L}_{\ell}^{-1}[\calP_{k}(x,z,t)],
  \end{equation}
  where ${\mathbb L}_{\ell}^{-1}$ denotes the inverse discrete Laplace transform. 
 Finally, note that given a discrete Laplace transform $\widetilde{f}(z)$, the inverse transform is defined by a contour integral on the unit circle $C$:
  \begin{equation}
  f(\ell)={\mathbb L}_{\ell}^{-1} [\widetilde{f}]=\oint_C\frac{\widetilde{f}(z)}{2\pi i z^{\ell+1}}dz.
  \end{equation}
  For relatively simple transforms $\widetilde{f}(z)$, we can extract the inverse transform by expanding $\widetilde{f}(z)$ as a geometric series in $z$.

\subsection{Stochastic resetting}
\begin{figure}[t!]
  \centering
   \includegraphics[width=9cm]{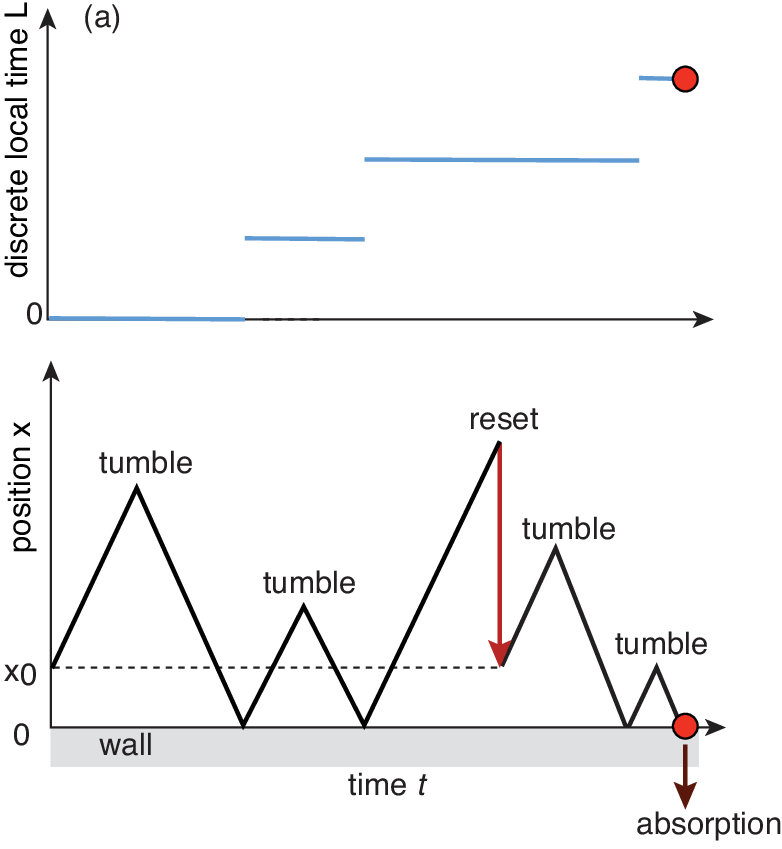}
  \caption{Sample trajectory of an RTP in the half-line with a partially absorbing wall at $x=0$ and subject to position and velocity resetting with $\sigma_0=1$. The corresponding jumps in the discrete local time are also shown. The local time does not reset.}
  \label{fig1}
  \end{figure}

So far we have considered two distinct Markov chains, the Poisson process that generates velocity reversals and a (non-Poissonian) counting process that keeps track of the number of wall collisions. Within the same spirit, stochastic resetting can be incorporated into the model by introducing a second independent Poisson process $\overline{N}(t)$, which specifies the number of resets occurring in the time interval $[0,t]$. That is,
 \begin{eqnarray}
\overline{N}(t)=n,\quad \overline{\calT}_{n}\leq t <\overline{\calT}_{n+1},\quad \P[\overline{N}(t)=n]=  \frac{(r t)^{n}\e^{-r t}}{n!},
\label{Pois2}
\end{eqnarray}
where $\overline{\calT}_n$ , $n\geq 1$, denotes the $n$th time that the RTP resets. The corresponding inter reset times $\overline{\tau}_n =\overline{\calT}_n-\overline{\calT}_{n-1}$ are exponentially distributed with
\begin{equation}
 \P[\overline{\tau}_n\in [s,s+ds]]=r \e^{-r s}ds .
\end{equation} 
Suppose that both $X(t)$ and $\sigma(t)$ instantaneously reset to their initial values $(x_0,\sigma_0)$, whereas $L(t)$ is not changed.  This scenario is illustrated in Fig. \ref{fig1} for $\sigma_0=1$, which shows a sample trajectory consisting of four tumbling events, one reset and three reflections prior to absorption. The jumps in the discrete local time are also shown.

Positional resetting can be implemented as a jump-diffusion process with
\begin{eqnarray}
 X(t)&=x_0+\int_0^tv\sigma(s)ds+\sum_{n=1}^{\overline{N}(t)}  [x_0-X(\overline{\calT}_n^-)]\nonumber \\
 &=x_0+\int_0^t\left [v\sigma(s)ds +[x_0-X( s^-)]d\overline{N}(s)\right ]
\label{intXr}
\end{eqnarray}
and
\begin{eqnarray}
\label{dNdefr}
d\overline{N}(t) =\overline{h}(t)dt,\quad \overline{h}(t)=\sum_{n=1}^{\infty}\delta(t-\overline{\calT}_{n})  .
\end{eqnarray}
Setting $t=\overline{\calT}_n$ and $t=\overline{\calT}_n^-$ in equation (\ref{intXr}) and subtracting the resulting pair of equations shows that $X(\overline{\calT}_n)= X(\overline{\calT}_{n}^-)+x_{0}- X(\overline{\calT}_{n}^-)=x_0$
which represents to instantaneous resetting. Resetting of the velocity state can be dealt with in a similar fashion. Equations (\ref{PDMPa}) and (\ref{PDMPb}) thus become
\numparts
\begin{eqnarray}
\label{rPDMPa}
dX(t)&=v\sigma(t)dt+(x_0-X(t^-))d\overline{N}(t),\\
 d\sigma(t)&= [\sigma_0-\sigma(t^-)]d\overline{N}(t)-2\sigma(t^-)[dN(t)+ dL(t)],
 \label{rPDMPb}\\
dL(t)&= \eta(t) dt,
 \label{rPDMPc}
 \end{eqnarray}
 \endnumparts
 together with the stopping condition (\ref{TA}).

We can extend the derivation of the discrete local time propagator with position and local time resetting as follows. First, equation (\ref{V1}) has an additional term on the right-hand side:
\begin{eqnarray}
\fl &\bigg [f(x_0,L(t),\sigma_0)-f(X(t^-),L(t^-),\sigma(t^-))\bigg ] d\overline{N}(t)\nonumber \\
\fl &\quad =\overline{h}(t)dt \sum_{\ell \geq 0} \sum_{k=\pm 1}  \int_0^{\infty}  [\delta(x-x_0)\delta_{k,\sigma_0}\delta_{\ell,L(t)}- \rho_{k,\ell}(x,t^-) ] f(x,\ell,k)  dx.
\end{eqnarray}
It follows that the SPDE (\ref{spuda}) for $\rho_k$ becomes
\begin{eqnarray}
  \frac{\partial \rho_{k,\ell}}{\partial t}&= -vk\frac{\partial}{\partial x} \rho_{k,\ell}(x, t)+h(t) [\rho_{-k,\ell}(x, t^-)-\rho_{k,\ell}(x, t^-)]
 \nonumber \\
  &\quad  +\overline{h}(t)\left  [\delta(x-x_0)\delta_{k,\sigma_0}\int_0^{\infty}\rho_{\ell}(y,t)dy- \rho_{k,\ell}(x,t^-) \right ] ,
  \end{eqnarray}
supplemented by the boundary condition (\ref{spudb}). We have set $\rho_{\ell}=\rho_{1,\ell}+\rho_{-1,\ell}$. Averaging with respect to both Poisson processes leads to the CK equations 
\numparts
\begin{eqnarray}
\label{rCKpropa}
& \frac{\partial P_{k,\ell}(x,t)}{\partial t}=-vk\frac{\partial P_{k,\ell}(x,t)}{\partial x}+\alpha [P_{-k,\ell}(x,t)-P_{k,\ell}(x,t)]\nonumber \\
  & \hspace{2cm} +r[\delta(x-x_0)\delta_{k,\sigma_0}S_{\ell}(t)-P_{k,\ell}(x,t) ],
\\
 &P_{1,\ell+1}(0,t)=P_{-1,\ell}(0,t),\quad \ell \geq 0.
 \label{rCKpropb}
\end{eqnarray}
\endnumparts
for $\ell\geq 0, \ k=\pm 1,\ x \in \R$, where $S_{\ell}$ is the marginal probability
\begin{equation}
S_{\ell}(t):=\int_0^{\infty}\E[ \rho_{\ell}(x,t)]dx.=\int_0^{\infty} [P_{1,\ell}(x,t)+P_{-1,\ell}(x,t)]dx.
\end{equation}
Note that $S_{\ell}(0)=\delta_{\ell,0}$ and
\begin{eqnarray}
\fl \frac{dS_{\ell}(t)}{dt}&=\int_0^{\infty}\left[\frac{\partial P_{1,\ell}(x,t)}{\partial t}+\frac{\partial P_{-1,\ell}(x,t)]}{\partial t} \right ]dx\nonumber \\
\fl &=\int_0^{\infty}\bigg [-v\frac{\partial P_{1,\ell}(x,t)}{\partial x}+v\frac{\partial P_{-1,\ell}(x,t)}{\partial x} \nonumber \\
\fl &\hspace{2cm}  +r\bigg (\delta(x-x_0) S_{\ell}(t)-P_{1,\ell}(x,t) -P_{-1,\ell}(x,t) \bigg )\bigg ]dx\nonumber \\
\fl &=v[P_{1,\ell}(0,t)-P_{-1,\ell}(0,t)]\equiv J_{\ell}(0,t).
\label{cool}
\end{eqnarray}

\begin{figure}[t!]
  \centering
   \includegraphics[width=9cm]{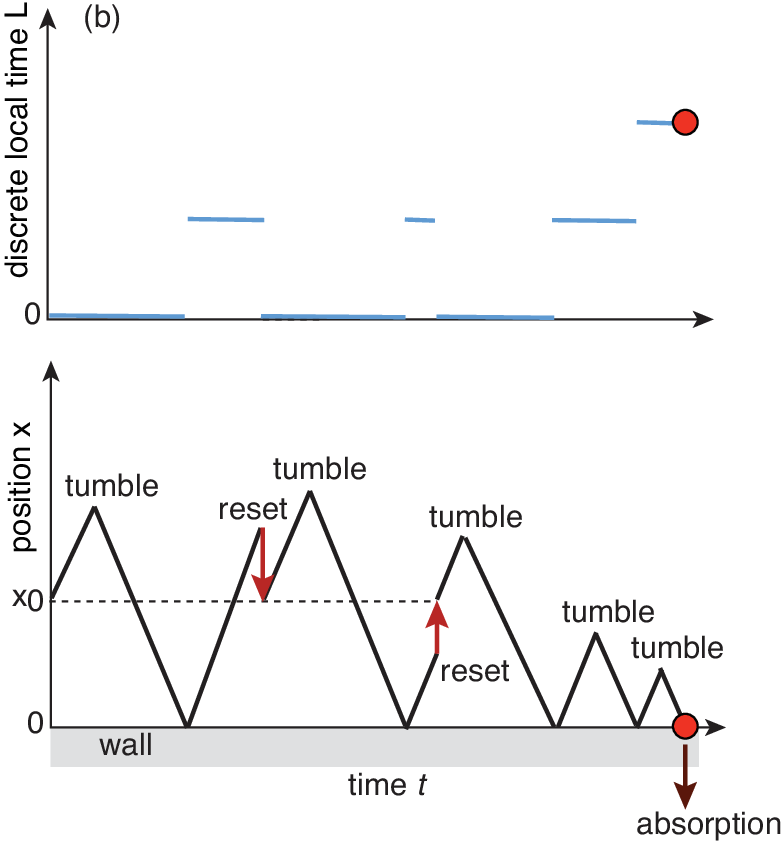}
  \caption{Sample trajectory of an RTP in the half-line with a partially absorbing wall at $x=0$ and subject to position, velocity and local time resetting with $\sigma_0=1$. The corresponding jumps in the discrete local time are also shown.}
  \label{fig2}
  \end{figure}

As previously highlighted for encounter-based models of diffusion with resetting \cite{Bressloff22b}, the assumption that the discrete local time does not reset is physically motivated by the idea that when a particle hits a wall it can alter the reactivity of an absorbing substrate. In this case one would not expect the local time to reset when the particle position resets. An alternative scenario involves the substrate modifying a discrete internal state $\calU(t)$ of the particle that affects the probability of absorption. More specifically, $\calU(t)=F(L(t))$ with $F(0)=0$ and $F'(\ell)> 0$ for all $\ell\geq 0$. Hence, $\calU(t)$ is a strictly monotonically increasing function of the local time. The stopping condition for absorption is then taken to be of the form
\begin{equation}
\label{TU}
{\mathcal T}=\inf\{t>0:\ \calU(t) >\widehat{\calU}\},
\end{equation}
where $\widehat{\calU}$ is a random variable with probability density $p(u)$. Since $F(a)$ is strictly monotonic, we have the equivalent stopping condition
\begin{equation}
\label{TA2}
{\mathcal T}=\inf\{t>0:L(t) >\widehat{\ell}=F^{-1}(\widehat{\calU})\}.
\end{equation}
One major difference from the substrate modification scenario is that one can now plausibly assume that $\calU(t)$ and thus $L(t)$ also reset:
$(X(t),L(t),\sigma(t))\rightarrow (X_0,0,\sigma_0)$. This is illustrated in Fig. \ref{fig2}. \pcb{(We also assume that a new random threshold $\widehat{\ell}$ is generated following each resetting.)} From a mathematical perspective, allowing the local time to reset means that propagator resetting can be treated as a renewal process. On the other hand, since the local time $L(t)$ is no longer a monotonically increasing function of $t$, the relationship (\ref{bob}) between $p^{\Psi}$ and the propagator breaks down. In the following section we assume that $L(t)$ does not reset. The alternative scenario will be considered in section 5.

 \setcounter{equation}{0}
\section{MFPT for generalized absorption without local time resetting} 

Given the local time propagators $P_{k,l}(x,t)$ for fixed initial (and resetting) data $(x_0,\sigma_0)$ and the associated marginal probability $S_{\ell}(t)$, let $p^{\Psi}(x,t)$ be the probability density for a partially absorbing wall with a prescribed local time threshold function $\Psi(\ell)$. The corresponding survival probability is
\begin{equation}
\label{SP}
S^{\Psi}(t)=\int_0^{\infty}p^{\Psi}(x,t)dx=\sum_{\ell=0}^{\infty}\Psi(\ell-1)S_{\ell}(t).
\end{equation}
If $T^{\Psi}$ denotes the first passage time for the given initial data, then ${\rm Prob}[T^{\Psi} \leq t] =1-S^{\Psi}(t)$. The FPT density $f^{\Psi}(t)=- {\partial S^{\Psi}(t)}/{\partial t}$ and the MFPT is 
\begin{eqnarray}
  \tau^{\Psi}&:=  \int_0^{\infty}t f^{\Psi}(t)dt =\int_0^{\infty}{S}^{\Psi}(t)dt\nonumber \\
&=\widetilde{S}^{\Psi}(s=0)=\sum_{\ell=0}^{\infty}\Psi(\ell-1)\widetilde{S}_{\ell}(s=0).
\end{eqnarray}
Note that higher-order moments of the FPT density can also be calculated from $\widetilde{S}_{\ell}(s)$. That is,
\begin{eqnarray}
\fl   \tau_n^{\Psi}&:=  \int_0^{\infty}t^n f^{\Psi}(t)dt =n\int_0^{\infty}t^{n-1}{S}^{\Psi}(t)dt\nonumber \\
\fl &=n\left (-\frac{d}{ds}\right )^{n-1}\left . \widetilde{S}^{\Psi}(s)\right |_{s=0}=n\sum_{\ell=0}^{\infty}\Psi(\ell-1)\left (-\frac{d}{ds}\right )^{n-1}\left .\widetilde{S}_{\ell} (s=0)\right |_{s=0}.
\end{eqnarray}

In order to determine $\widetilde{S}_{\ell}(s)$, we perform a double Laplace transform of the propagator equations (\ref{rCKpropa}) and (\ref{rCKpropb}) with respect to $t$ and $\ell$. Setting
\begin{eqnarray}
\label{dLP}
\calPP_{k}(x,z,s )&=\int_0^{\infty} dt\, \e^{-st}\sum_{\ell=0}^{\infty} z^{\ell} P_{k,\ell}(x,t),\quad 0\leq z \leq 1,
\end{eqnarray}
we obtain the propagator BVP 
\numparts
\begin{eqnarray}
\label{rLT1}
 & vk\frac{\partial \calPP_{k}(x,z,s)}{\partial x}+(s+\alpha+r) \calPP_{k}(x,z,s)\\
 &\quad =\alpha \calPP_{-k}(x,z,s) +\left (1+r\widetilde{\calS}(z,s)\right )\delta(x-x_0)\delta_{k,\sigma_0},\nonumber\\
 \fl &\calPP_{1}(0,z,s)= z\calPP_{-1}(0,z,s) ,
 \label{rLT2}
\end{eqnarray}
\endnumparts
with $x,x_0>0$. and
\begin{equation}
 \widetilde{\calS}(z,s)=\int_0^{\infty} dt\, \e^{-st}\sum_{\ell=0}^{\infty} z^{\ell} S_{\ell}(t)dt.
\end{equation}
Performing a double Laplace transform of equation (\ref{cool}) shows that
\begin{eqnarray}
s \widetilde{\calS}(z,s)-1=v[\calPP_{1}(0,z,s)-\calPP_{-1}(0,z,s)].
\end{eqnarray}
Imposing the boundary condition yields
\begin{eqnarray}
s \widetilde{\calS}(z,s)=1-(1-z)v \calPP_{-1}(0,z,s).
\label{tilS}
\end{eqnarray}
Given the double Laplace transform $\widetilde{\calS}(z,s)$, the MFPT for the 
 general distribution $\Psi$ can be expressed as
\begin{eqnarray}
  \tau^{\Psi}&=\sum_{\ell=0}^{\infty}\Psi(\ell-1){\mathbb L}_{\ell}^{-1}[\widetilde{\calS}(0,z)].
  \label{tPsi}
\end{eqnarray}
This reduces to
\begin{equation}
\tau=\widetilde{\calS}(0,z_0)
 \label{0tPsi}
\end{equation}
for the geometric distribution $\Psi(\ell)=z_0^{\ell}$.

\subsection{MFPT $\tau^{\Psi}$ for $\sigma_0=\pm 1$} The next step is to solve equations (\ref{rLT1}) and (\ref{rLT2}). Since both the boundary condition and the resetting protocol differ from the examples considered in Ref. \cite{Evans18}, we present the details of the calculation in appendix A. In addition, given that the velocity resetting protocol is not symmetric, we consider the cases $\sigma_0=\pm 1$ separately. In the case $\sigma_0=1$, we find that
\begin{eqnarray}
\lim_{s\rightarrow 0} \widetilde{\calS}(z,s)=-\frac{1}{r}+\frac{  \e^{\Lambda_0(r) x_0}}{r} \frac{1}{1-z}\left [\frac{1- z\Gamma_-(r))}{\Gamma_-(r)}\right ] ,
\label{A0}
\end{eqnarray}
 with
 \begin{equation}
 \label{gam}
\Gamma_{\pm}(r):=\frac{1}{\alpha}\bigg [r+\alpha\pm \sqrt{r(2\alpha+r )}\bigg ]
\end{equation}
such that
\begin{equation}
\Gamma_-(r)\in [0,1],\quad \frac{1}{\Gamma_+(r)} \in [0.1].
\end{equation}
First suppose that the partially absorbing boundary has a constant rate of absorption $\kappa_0$, see the Robin boundary condition (\ref{calJ}).
Setting $z=z(\kappa_0)=v/(v+\kappa_0)$ in equation (\ref{0tPsi}) we have the MFPT
\begin{eqnarray}
\tau_{\sigma_0=1}&=\lim_{s\rightarrow 0} \widetilde{\calS}(z(\kappa_0),s)\nonumber \\
& =-\frac{1}{r}+\frac{  \e^{\Lambda_0(r) x_0}}{r} \frac{1}{\kappa_0}\left [\frac{v+\kappa_0-v \Gamma_-(r)}{\Gamma_-(r)} \right ].
\label{tau0p}
\end{eqnarray}
In the limit $\kappa_0\rightarrow 0$ the wall becomes totally reflecting and $\tau_{\sigma_0=1}\rightarrow \infty$. On the other hand, in the case of a totally absorbing wall, we have
\begin{equation}
\lim_{\kappa_0\rightarrow \infty} \tau_{\sigma_0=1} =--\frac{1}{r}+\frac{\alpha}{r}\left [\frac{ \e^{\Lambda_0(r)x_0}}{r+\alpha-\sqrt{r(2\alpha+r )}}\right ].
\end{equation} 
Note that this result differs from the corresponding MFPT for velocity randomization, see equation (\ref{sym0}).
In order to consider a non-geometric threshold distribution $\Psi(\ell)$, we need to invert equation (\ref{A0}) with respect to $z$:
\begin{eqnarray}
 \lim_{s\rightarrow 0} \widetilde{S}_{\ell}(s)&=\oint_C\frac{1}{2\pi i z^{\ell+1}}\left [-\frac{1}{r}+\frac{  \e^{\Lambda_0(r) x_0}}{r} \frac{1}{1-z}\left (\frac{1-z\Gamma_-(r)}{\Gamma_-(r)}\right )
\right ]dz\nonumber \\
   &=-\frac{\delta_{\ell,0}}{r}+\frac{  \e^{\Lambda_0(r) x_0}}{r}  \left [\frac{1-\Gamma_-(r)(1-\delta_{\ell,0})}{\Gamma_-(r)}\right ].
\end{eqnarray}
Hence,
\begin{eqnarray}
 \tau_{\sigma_0=1}^{\Psi}&= \lim_{s\rightarrow 0}\sum_{\ell=0}^{\infty}\Psi(\ell-1) \widetilde{S}^{\Psi}_{\ell}(s) \\
  &=-\frac{1}{r}+\frac{  \e^{\Lambda_0(r) x_0}}{r}  \left [\frac{ [1-\Gamma_-(r)] \sum_{\ell=0}^{\infty}\Psi(\ell-1)+1 }{\Gamma_-(r)}\right ].
\label{tauP1}
\end{eqnarray}
(Here and in the following we assume that the sum with respect to $\ell$ is uniformly convergent so that we can swap the order of summation and the small-$s$ limit.)

Proceeding along similar lines in the case $\sigma_0=-1$ we find that (see appendix A) 
\begin{eqnarray}
\lim_{s\rightarrow 0} \widetilde{\calS}(z,s)=-\frac{1}{r}+\frac{  \e^{\Lambda_0(r) x_0}}{r} \frac{1}{1-z}\frac{\Gamma_+(r)-z }{\Gamma_+(r)} ,
\label{B0}
\end{eqnarray}
with $\Gamma_+(r)$ defined in equation (\ref{gam}).
Setting $z=z(\kappa_0)=v/(v+\kappa_0)$ in equation (\ref{0tPsi}) now gives 
\begin{eqnarray}
 \tau_{\sigma_0=-1}&=\lim_{s\rightarrow 0} \widetilde{\calS}(z(\kappa_0),s)\nonumber \\
 &=-\frac{1}{r}+\frac{  \e^{\Lambda_0(r) x_0}}{r} \frac{1}{\kappa_0}\left [\frac{(v+\kappa_0)\Gamma_+(r)-  v }{\Gamma_+(r)} \right ].
\label{tau0m}
\end{eqnarray}
In particular,
\begin{equation}
\lim_{\kappa_0\rightarrow \infty} \tau_{\sigma_0=-1} =-\frac{1}{r}+\frac{\e^{\Lambda_0(r)x_0}}{r}.
\label{bon}
\end{equation} 
On the other hand, for a non-geometric threshold distribution $\Psi(\ell)$, we invert equation (\ref{B0}) with respect to $z$ to give
\begin{eqnarray}
 \lim_{s\rightarrow 0} \widetilde{S}_{\ell}(s)&=\oint_C\frac{1}{2\pi i z^{\ell+1}}\left [-\frac{1}{r}+\frac{  \e^{\Lambda_0(r) x_0}}{r} \frac{1}{1-z}\left [\frac{\Gamma_+(r)-z }{\Gamma_+(r)}\right ] 
\right ]dz\nonumber \\
   &=-\frac{\delta_{\ell,0}}{r}+\frac{  \e^{\Lambda_0(r) x_0}}{r}  \left [\frac{\Gamma_+(r)- (1-\delta_{\ell,0})}{v(\lambda_0(r) +\Lambda)_0(r)}\right ].
\end{eqnarray}
Equation (\ref{tPsi}) then shows that
\begin{eqnarray}
\fl  \tau_{\sigma_0=-1}^{\Psi}&= \lim_{s\rightarrow 0} \widetilde{S}^{\Psi}_{\ell}(s) \nonumber \\
\fl   &=-\frac{1}{r}+\frac{  \e^{\Lambda_0(r) x_0}}{r}  \left [\frac{[\Gamma_+(r)-1]\sum_{\ell=0}^{\infty}\Psi(\ell-1)+\Gamma_+(r)}{\Gamma_+(r)}\right ].
\label{tauP2}
\end{eqnarray}

 \subsection{Parameter dependence of the MFPT}

 Equations (\ref{tauP1}) and (\ref{tauP2}) determine the MFPT for $\sigma_0=\pm 1$ and a general distribution $\Psi(\ell)$.
It can be seen that both of these equations involve the discrete Laplace transform of $\Psi(\ell)$, defined according to
 \begin{equation}
 \widetilde{\Psi}(z)=\sum_{\ell=0}^{\infty} z^{\ell}\Psi(\ell-1),\quad \Psi(-1)=1.
 \end{equation}
 Equation (\ref{psi}) implies that
\begin{eqnarray}
\widetilde{\psi}(z)&\equiv \sum_{\ell=0}^{\infty} z^{\ell}\psi(\ell)\nonumber \\
&=\sum_{\ell=0}^{\infty} z^{\ell}[\Psi(\ell-1)-\Psi(\ell)]\nonumber \\
&=\widetilde{\Psi}(z)\left (1-\frac{1}{z}\right )+\frac{1}{z},
\end{eqnarray}
so that
\begin{equation}
\widetilde{\Psi}(z)=\frac{\widetilde{\psi}(z)-1/z}{1-1/z}.
\label{Ppsi}
\end{equation}
Since $\widetilde{\psi}(1)=1$, we can use the Leibniz rule to deduce that
\begin{equation}
\sum_{\ell=0}^{\infty} \Psi(\ell-1)\equiv \lim_{z\rightarrow 1^-} \widetilde{\Psi}(z)= \widetilde{\psi}'(0)+1=\E[\ell]+1 .
\end{equation}
We have used the fact that $\widetilde{\psi}(z)$ is the moment generating function of the local time threshold.
Using these results, equations (\ref{tauP1}) and (\ref{tauP2}) become
\begin{eqnarray}
 \tau_{\sigma_0=1}^{\Psi}
  &=-\frac{1}{r}+\frac{  \e^{\Lambda_0(r) x_0}}{r}  \left [\frac{ 1+(1-\Gamma_-(r)) \E[\ell] }{\Gamma_-(r)}\right ],
\label{restauP1}
\end{eqnarray}
and
\begin{eqnarray}
 \tau_{\sigma_0=-1}^{\Psi} 
 &=-\frac{1}{r}+\frac{  \e^{\Lambda_0(r) x_0}}{r}  \left [\frac{[\Gamma_+(r)-1]\E[\ell]+\Gamma_+(r)}{\Gamma_+(r)}\right ].
\label{restauP2}
\end{eqnarray}
We conclude that the only dependence of the MFPT on the local time threshold distribution $\Psi(\ell)$ is via the mean $\E[\ell]$, assuming the latter exists. 
It follows that the same qualitative behavior is obtained irrespective of whether absorption is Markovian or non-Markovian. Indeed, we recover equations (\ref{tau0p}) and (\ref{tau0m}) on setting $\E[\ell]=v/\kappa_0$, which is the mean of the distribution $\psi(\ell)=(1-z_0)z_0^{\ell}$ for $z_0=v/(v+\kappa_0)$.

 \begin{figure}[t!]
  \centering
   \includegraphics[width=13cm]{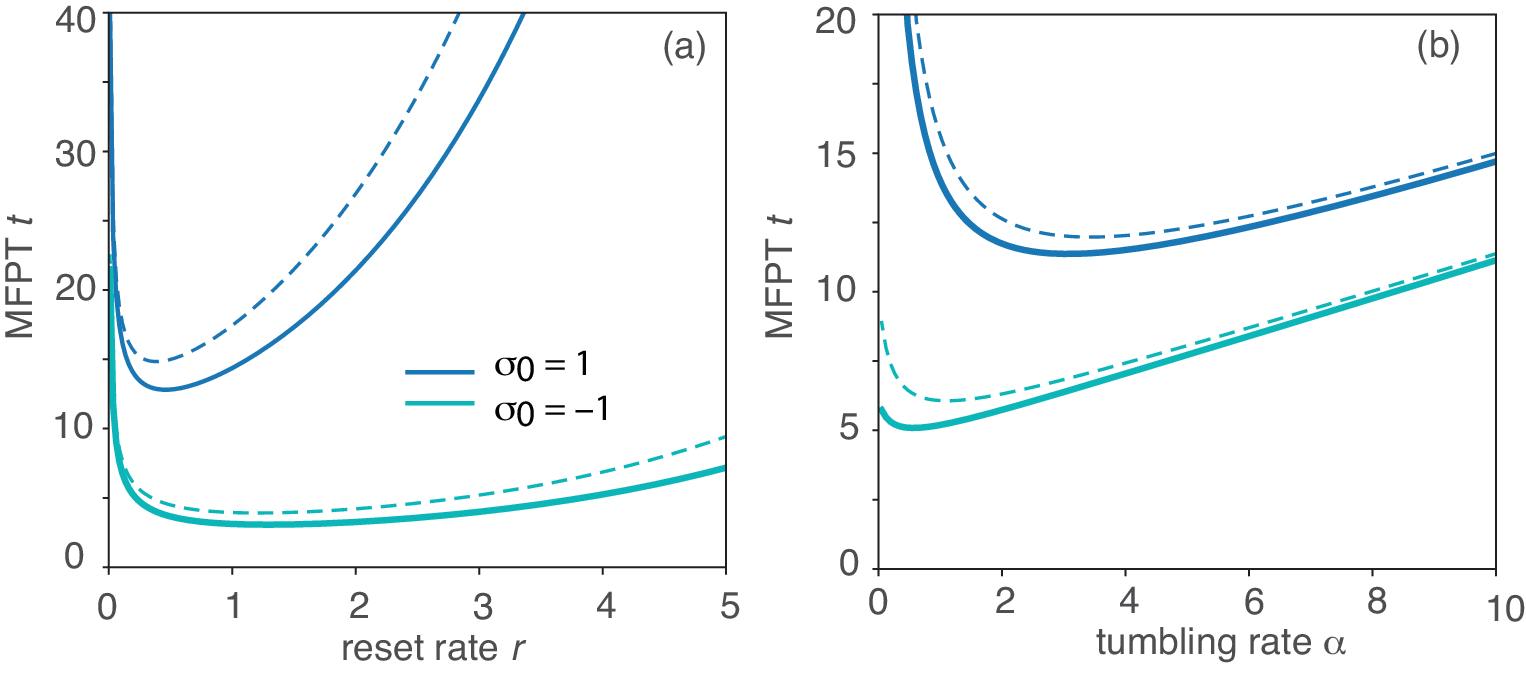}
  \caption{Plots of the MFPTs $\tau_{\sigma_0=\pm1}^{\Psi}$ given by equations (\ref{restauP1}) and (\ref{restauP2}), respectively, for $\E[\ell]=1$. (a) MFPT as a function of the resetting rate $r$ for $\alpha=1$. (b) MFPT as a function of the tumbling rate $\alpha$ for $r=0.2$. Other parameters are $v=1$, $\E[\ell]=1$ and $x_0=0.5$. (Dashed curves show the corresponding plots in the case of local time resetting and a Poisson distribution $\psi(\ell)$ with unit mean, see section 5.)}
  \label{fig3}
  \end{figure}
  
  \begin{figure}[t!]
  \centering
   \includegraphics[width=13cm]{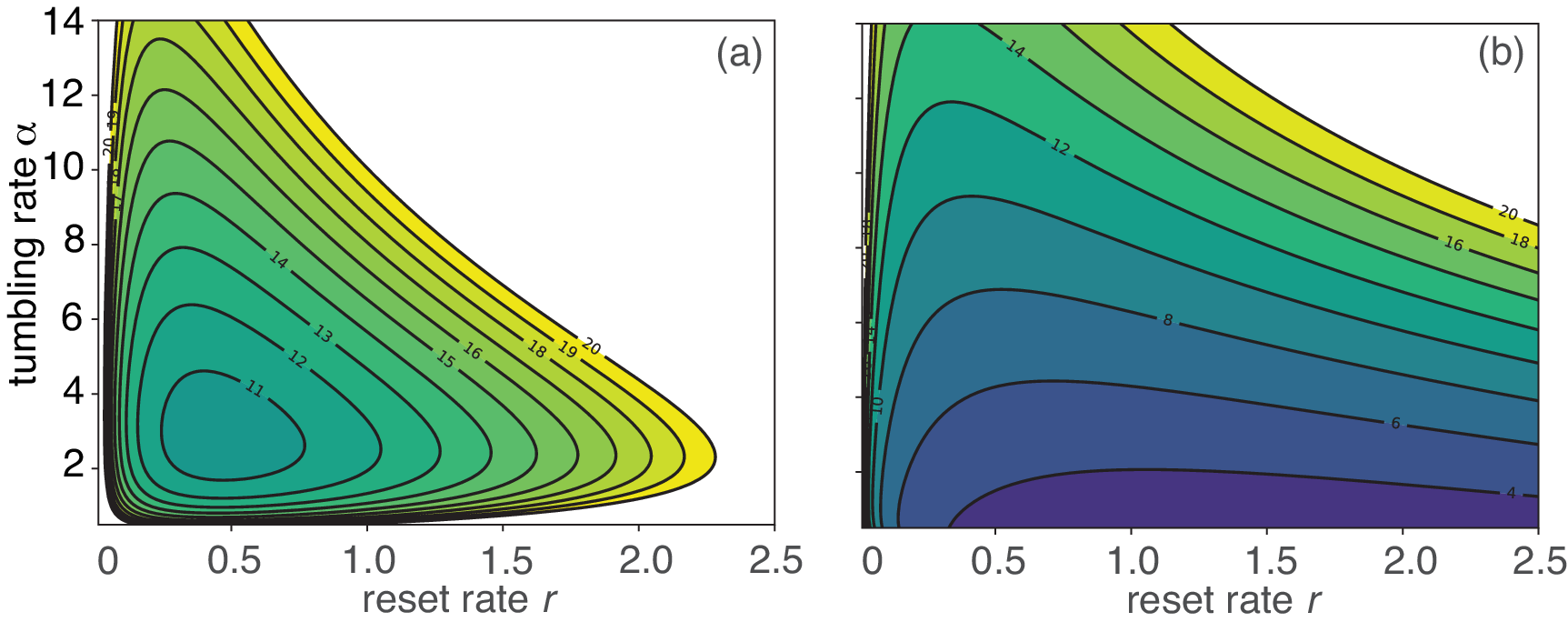}
  \caption{Contour plots of the MFPTs (a) $\tau_{\sigma_0=1}^{\Psi}$ and (b) $\tau_{\sigma_0=-1}^{\Psi}$ in the $(r,\alpha)$-plane. Other parameters are $v=1$ and $x_0=0.5$.}
  \label{fig4}
  \end{figure}
  
   \begin{figure}[t!]
  \centering
   \includegraphics[width=13cm]{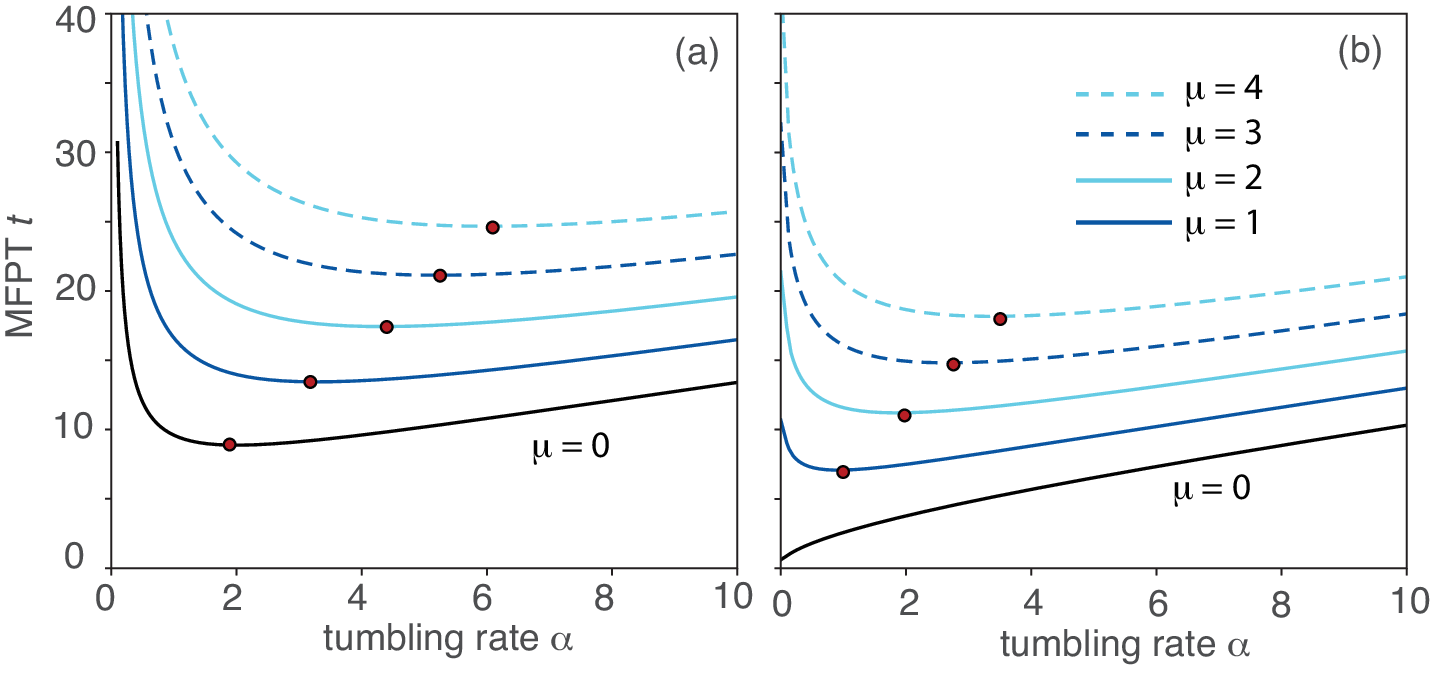}
  \caption{Plots of the MFPTs (a) $\tau_{\sigma_0=1}^{\Psi}$ and (b) $\tau_{\sigma_0=-1}^{\Psi}$ given by equations (\ref{restauP1}) and (\ref{restauP2}), , respectively, as a function of $\alpha$ and various mean thresholds $\mu=\E[\ell]$. Other parameters are $v=1$, $r=0.1$ and $x_0=0.5$.}
  \label{fig5}
  \end{figure}

Since the MFPTs are clearly monotonically increasing functions of $x_0$ and $\E[\ell]$, we focus on their dependence on $r$ and $\alpha$, see also equations (\ref{Lam}) and (\ref{gam}). In Fig. \ref{fig3}(a) we show some example plots of the MFPTs $\tau_{\sigma_0=\pm 1}^{\Psi} $ as functions of the resetting rate $r$ for fixed tumbling rate $\alpha>0$. We observe the usual unimodal variation with $r$ for search processes on the half-line. Moreover, $\tau_{\sigma_0=-1}^{\Psi} <\tau_{\sigma_0=1}^{\Psi} $ since the particle resets to the velocity state moving towards (away from) the wall when $\sigma_0=-1$ ($\sigma_0=+1$). We also find a non-monotonic dependence on $\alpha$ for fixed $r$ and $\sigma_0=1$ with $\lim_{\alpha \rightarrow 0}\tau_{\sigma_0=1}^{\Psi} =\infty$, see Fig. \ref{fig3}(b). That is, if the probability of tumbling approaches zero, then the particle remains in the right-moving state and never hits the wall. On the other hand, non-monotonicity only occurs for $\sigma_0=-1$ when $r$ is sufficiently small. (If resetting events are rare then a non-zero rate of tumbling is needed in order to reverse trajectories that reflect off the wall without being absorbed.) In addition the MFPT $\tau_{\sigma_0=-1}^{\Psi} $ is finite in the limit $\alpha \rightarrow 0$ with $r>0$. The dual dependence on $r$ and $\alpha$ is further illustrated by the contour plots in Fig. \ref{fig4}. In particular, Fig. \ref{fig4}(a) indicates that there exists a unique minimum in the $(r,\alpha)$-plane when $\sigma_0=1$. Moreover, we find that this minimum moves towards the origin as $x_0$ increases from zero (not shown). Finally, note that in the limit $\E[\ell]\rightarrow 0$ we recover a totally absorbing wall. In this case the MFPT for $\sigma_0=-1$ is a monotonically increasing function of $\alpha$, which follows from equation (\ref{bon}). The variation of the MFPTs with $\alpha$ for various values of $\E[\ell]$ are shown in Fig. \ref{fig5}.

\section{Renewal theory}

In the special case of a constant reactivity $\kappa_0$ (Robin boundary condition), the survival probability ${\mathcal S}(z,t)$ with $z=v/(v+\kappa_0)$ satisfies a renewal equation of the form (\ref{ren}). Within the encounter-based framework, this reflects the fact that for a geometric local time threshold distribution, the probability of absorption is memoryless. It follows that
\begin{equation}
\label{ren3}
\widetilde{\calS}(z,s)=-\frac{1}{r} +\frac{1}{r}\frac{1}{1-r\widetilde{\calS}_0(z,s+r)}
\end{equation}
This result is independent of whether or not the local time resets. \pcb{One way to understand this result is as follows. Suppose that the particle is absorbed at time $t$ following the  $M$th collision and let $X([0,t])$ denote the given sample path. If there is a single local time reset at time $\tau \in (0,t)$ then the probability of absorption is obtained by averaging with respect to the local time threshold (which is also reset):
\[\fl \E[L_1(\tau) <\widehat{\ell}_1,L_2(t-\tau)=\widehat{\ell}_2]=z_0^{L_1(\tau)-1}(1-z_0)z_0^{L_2(t-\tau)}=(1-z) z^{M-1}
\]
with $L_1(0)=L_2(\tau)=0$ and $L_1(\tau)+L_2(t-\tau)=M$. That is, the probability of absorption given a trajectory $X([0,t])$ just depends on the number of collisions, which is independent of whether or not the local time resets. (This argument breaks down for non-geometric distributions.)
}

In the case of a non-geometric distribution $\Psi$ and assuming the local time does not reset, we have
\begin{eqnarray}
\widetilde{S}^{\Psi}(s)&\equiv \sum_{\ell=0}^{\infty} \Psi(\ell-1) {\mathbb L}^{-1}_{\ell}[\widetilde{\calS}(z,s)] \neq  -\frac{1}{r} +\frac{1}{r}\frac{1}{1-r\widetilde{\calS}^{\Psi}_0(s+r)}.
\end{eqnarray}
That is, the survival probability for a non-geometric distribution $\Psi$ does not satisfy a renewal equation.
Now suppose that the local time does reset to zero as illustrated in Fig. \ref{fig2}. The propagator equations (\ref{rCKpropa}) and (\ref{rCKpropb}) become
\numparts
\begin{eqnarray}
\label{LCKpropa}
& \frac{\partial P_{k,\ell}(x,t)}{\partial t}=-vk\frac{\partial P_{k,\ell}(x,t)}{\partial x}+\alpha [P_{-k,\ell}(x,t)-P_{k,\ell}(x,t)]\nonumber \\
  & \hspace{2cm} +r[\delta(x-x_0)\delta_{k,\sigma_0}\delta_{\ell,0}-P_{k,\ell}(x,t) ],
\\
 &P_{1,\ell+1}(0,t)=P_{-1,\ell}(0,t),\quad \ell \geq 0.
 \label{LCKpropb}
\end{eqnarray}
\endnumparts
In this case the local time propagator satisfies a first renewal equation of the form
\begin{equation}
 P_{k,\ell}(x,t)= \e^{-rt}P^{(0)}_{k,\ell}((x,t)+r\int_0^t\e^{-r \tau}P_{k,\ell}(x,t-\tau)d\tau .
\end{equation}
where $P^{(0)}$ is the propagator in the absence of resetting.
The first term on the right-hand side represents all trajectories that do not undergo any resettings, which occurs with probability $\e^{-rt}$. The second term represents the complementary set of trajectories that reset at least once with the first reset occurring at time $\tau$. Laplace transforming the renewal equation and rearranging shows that
\begin{equation}
\label{con}
\widetilde{P}_{k,\ell}(x,s)=\left (1+\frac{r}{s}\right ) \widetilde{P}^{(0)}_{k,\ell}(x,s).
\end{equation}
However, in contrast to position and velocity resetting alone, one cannot simply take
$p_k^{\Psi}(x,t)=\sum_{\ell=0}^{\infty}\Psi(\ell-1)P_{k,\ell}(x,t)$, since $L(t)$ is no longer a monotonically increasing function of time $t$. Therefore, following our previous analysis of diffusion processes with resetting \cite{Bressloff22b}, we partition the set of contributing paths according to the number of resettings and, for a given number of resettings, decomposing the path into time intervals over which $L(t)$ is monotonically increasing. Let ${\mathcal I}_t$ denote the number of resettings in the interval $[0,t]$ and let ${\mathcal T}=\inf\{t>0, L(t) >\widehat{\ell}\}$. Then
\begin{eqnarray}
\fl p^{\Psi}(x,t)d\x&=\e^{-rt}\P[X(t)\in [x,x+dx]|X(0)=x_0,\, {\mathcal T}>t,\, {\mathcal I}_t=0]\\
\fl &\quad +r\e^{-rt}\P[X(t) \in [x,x+dx]|X(0)=x_0,\, {\mathcal T}>t,\, {\mathcal I}_t=1]\nonumber \\
\fl &\quad +r^2\e^{-rt}\P[X(t)\in [x,x+dx]|X(0)=x_0,\, {\mathcal T}>t,\, {\mathcal I}_t=2]+\ldots \nonumber 
\end{eqnarray}
That is,
\begin{eqnarray}
\fl p^{\Psi}(x,t)&=\e^{-rt} p_0^{\Psi}(x,t)+r\e^{-rt} \int_0^t p_0^{\Psi}(x,\tau)S_0^{\Psi}(t-\tau)d\tau\\
\fl &\quad +r^2\e^{-rt} \int_0^t \int_0^{t-\tau}p_0^{\Psi}(x,\tau)S_0^{\Psi}(t-\tau)S_0^{\Psi}(t-\tau-\tau')d\tau'd\tau+\ldots,\nonumber
\end{eqnarray}
where $S_0^{\Psi}$ is the survival probability without resetting. Laplace transforming the above equation and using the convolution theorem shows that
\begin{eqnarray}
 \p^{\Psi}(x,s)&=\p_0^{\Psi}(x,r+s)+r\p_0^{\Psi}(x,r+s)\widetilde{S}_0^{\Psi}(r+s)\nonumber \\
&\quad +r^2\p_0(x,r+s)\widetilde{S}_0^{\Psi}(r+s)^2+\ldots
\end{eqnarray}
Summing the geometric series thus yields the result
\begin{equation}
\label{prQ}
\p^{\Psi}(x,s)=\frac{\p_0^{\Psi}(x,r+s)}{1-r\widetilde{S}^{\Psi}_0(r+s)}.
\end{equation}
Finally, integrating both sides with respect to $x$ yields 
\begin{eqnarray}
\widetilde{S}^{\Psi}(s)&= \frac{\widetilde{S}^{\Psi}_0(s+r)}{1-r\widetilde{S}^{\Psi}_0(s+r)}=  -\frac{1}{r} +\frac{1}{r}\frac{1}{1-r\widetilde{S}^{\Psi}_0(s+r)}
\end{eqnarray}
 for general $\Psi$. In particular,
 \begin{eqnarray}
\tau^{\Psi}(s)&= -\frac{1}{r} +\frac{1}{r}\frac{1}{1-r\widetilde{S}^{\Psi}_0(r)}.
\label{tarp}
\end{eqnarray}
 
 We can determine $\widetilde{S}^{\Psi}_0(s)$ from the analysis of section 4 and appendix A. First suppose that $\sigma_0=1$. Setting $r=0$ in equation (\ref{A}) gives
 \begin{eqnarray}
  \frac{\alpha}{v}A_0(z,s)\left [   z-\frac{1}{\Gamma_-(s) }
\right ]=-\frac{\alpha}{v^2}   \e^{-\Lambda_0(s) x_0},
\label{Ar0}
\end{eqnarray}
with $\Gamma_-$ defined in equation (\ref{gam}). Moreover, equation (\ref{AA}) implies that
\begin{eqnarray}
s \widetilde{\calS}_0(z,s)=1-(1-z)vA_0(z,s).
\end{eqnarray}
Hence,
\begin{eqnarray}
 \widetilde{\calS}_0(z,s)=\frac{1}{s}-\frac{1-z}{s }\left [\frac{\Gamma_-(s)\e^{-\Lambda_0(s) x_0} }{1- z\Gamma_-(s)}\right ].
 \end{eqnarray}
 Inverting the discrete Laplace transform in $z$ then yields
\begin{eqnarray}
 \fl \widetilde{S}_{\ell,0}(s)=\frac{\delta_{\ell,0}}{s}-\frac{\Gamma_-(s)\e^{-\Lambda_0(s) x_0} }{s}\left [\delta_{\ell,0}-(1-\Gamma_-(s))\sum_{n\geq 1}\Gamma_-(s)^{n-1}\delta_{\ell,n}\right ],
 \end{eqnarray}
 and so
 \begin{eqnarray}
 \fl \widetilde{S}_{0}^{\Psi}(s)&=\frac{1}{s}-\frac{\Gamma_-(s)\e^{-\Lambda_0(s) x_0} }{s}\left [\frac{1}{\Gamma_-(s)}-\left (\frac{1}{\Gamma_-(s)}-1\right )\sum_{\ell\geq 0}\Gamma_-(s)^{\ell}\Psi(\ell-1)\right ]\nonumber\\
 \fl &=\frac{1}{s}-\frac{ \e^{-\Lambda_0(s) x_0} }{s}\left [1-\left (1-\Gamma_-(s)\right ) \widetilde{\Psi}(\Gamma_-(s))\right ].
 \label{SP1}
 \end{eqnarray}
 Similarly setting $r=0$ in equation (\ref{B}) with $\sigma_0=-1$ gives
\begin{eqnarray}
 \frac{\alpha}{v}\frac{B_0(z,s)}{z}\left [   1-\frac{z}{\Gamma_+(s) }
\right ]=\frac{\alpha}{v^2}  \e^{-\Lambda_0(s) x_0},
\label{B00}
\end{eqnarray}
with $\Gamma_+$ defined in equation (\ref{gam}) and (see equation (\ref{BB}))
\begin{eqnarray}
s \widetilde{\calS}_0(z,s)=1-\frac{1-z}{z}vB_0(z,s).
\end{eqnarray}
It follows that
\begin{eqnarray}
 \widetilde{\calS}_0(z,s)=\frac{1}{s}-\frac{1-z}{s }\left [\frac{\Gamma_+(s)\e^{-\Lambda_0(s) x_0} }{\Gamma_+(s)-z}\right ].
 \end{eqnarray}
 Inverting the discrete Laplace transform in $z$ then yields
\begin{eqnarray}
 \fl \widetilde{S}_{\ell,0}(s)=\frac{\delta_{\ell,0}}{s}-\frac{ \e^{-\Lambda_0(s) x_0} }{s}\left [\delta_{\ell,0}-(1-\Gamma_+(s)^{-1})\sum_{n\geq 1}\Gamma_+(s)^{1-n}\delta_{\ell,n}\right ],
 \end{eqnarray}
 and so
 \begin{eqnarray}
 \fl \widetilde{S}_{0}^{\Psi}(s)&=\frac{1}{s}-\frac{ \e^{-\Lambda_0(s) x_0} }{s}\left [\Gamma_+(s)-\left ( \Gamma_+(s)-1\right )\sum_{\ell\geq 0}\left (\frac{1}{\Gamma_+(s)}\right )^{\ell}\Psi(\ell-1)\right ]\nonumber\\
\fl & =\frac{1}{s}-\frac{ \e^{-\Lambda_0(s) x_0} }{s}\left [\Gamma_+(s)-\left ( \Gamma_+(s)-1\right )\widetilde{\Psi}(1/\Gamma_+(s))\right ].
 \label{SP2}
 \end{eqnarray}
 
 \begin{figure}[t!]
  \centering
   \includegraphics[width=13cm]{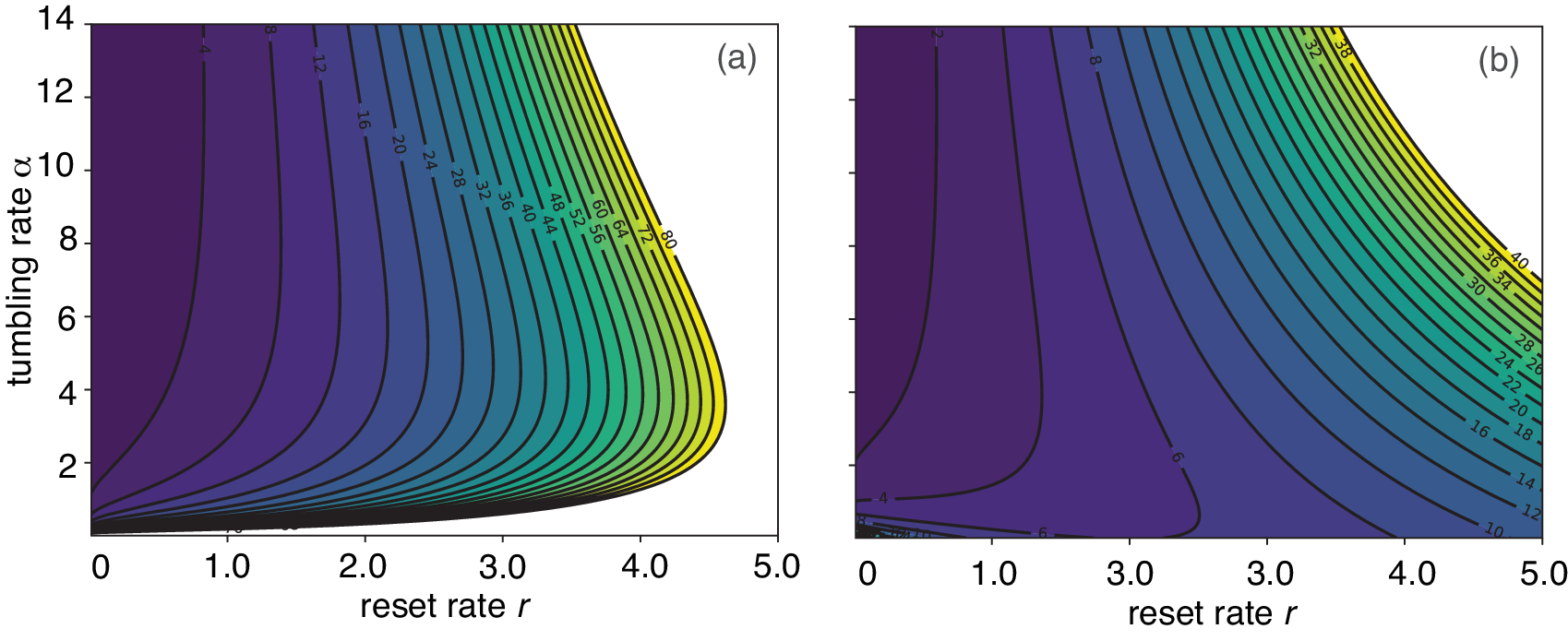}
  \caption{Contour plots of (a) $\Delta \tau_{\sigma_0=1}$ and (b) $\Delta \tau_{\sigma_0=-1}$ in the $(r,\alpha)$-plane. Other parameters are $v=1$, $\mu=1$ and $x_0=0.5$.}
  \label{fig6}
  \end{figure}
  
   \begin{figure}[b!]
  \centering
   \includegraphics[width=8cm]{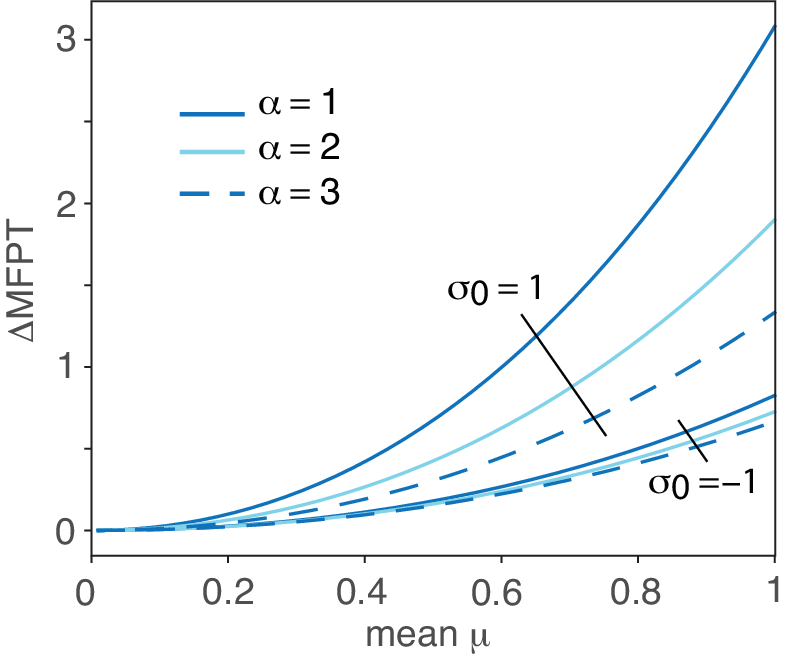}
  \caption{Plots of $\Delta \tau_{\sigma_0=1}$ and $\Delta \tau_{\sigma_0=-1}$ as a function of $\mu=\E[\ell]$ for different values of the tumbling rate $\alpha$. Other parameters are $v=1$, $r=1$ and $x_0=0.5$.}
  \label{fig7}
  \end{figure}

Equations (\ref{tarp}), (\ref{SP1}) and (\ref{SP2}) establish that in the case of local time resetting, the MFPT depends on the full statistics of the threshold distribution as expressed by the discrete Laplace transform $\widetilde{\Psi}(z)$, rather than just the mean threshold $\E[\ell]$.
 A simple example of a non-geometric distribution with finite moments is the Poisson distribution:
\begin{equation}
\label{Poiss}
\psi(\ell)=\frac{\mu^{\ell}\e^{-\mu}}{\ell!},\quad   \quad \widetilde{\psi}(z)=\e^{\mu(z-1)} .
\end{equation}
From equation (\ref{Ppsi}) we then have
\begin{equation}
 \widetilde{\Psi}(z)=\frac{\e^{\mu(z-1)}-1/z}{1-1/z},\quad \lim_{z\rightarrow 1^-} \widetilde{\Psi}(z)=1+\mu.
 \end{equation}
 Let
\begin{equation}
\Delta \tau_{\sigma_0} =\tau_{\sigma_0}^{\Psi_P(\mu)}-\tau_{\sigma_0}^{\Psi_G(\mu)},
 \end{equation}
 where $\Psi_P(\mu)$ and $\Psi_G(\mu)$ denote the Poisson and geometric distributions, respectively, with $\E[\ell]=\mu$. Note that $\Delta \tau_{\sigma_0} >0$ since local time resetting means that the expected time for $L(t)$ to cross the threshold $\widehat{\ell}$ is larger than when $L(t)$ does not reset, see the dashed curves in Fig. \ref{fig3}. In Fig. \ref{fig6} we show example contour plots of $\Delta \tau_{\pm 1} $ for fixed $\mu=1$. The non-monotonic dependence on $r$ and $\alpha$ can clearly be seen. We also find that the difference between the MFPTs with and without resetting increases monotonically with the mean $\mu$, as shown in Fig. \ref{fig7}.

 \section{Conclusion}  
 
 In this paper we combined our previous work on an encounter-based model of diffusion-mediated surface reactions with resetting \cite{Bressloff22b} and an encounter-based model of an RTP without resetting \cite{Bressloff22d}. The model of an RTP involves two major differences compared to a Brownian particle. First, the contact time $L(t) $ is a discrete random variable that counts the number of collisions with the (non-sticky) boundary, rather than a Brownian local time. In the case of the half-line, the latter is defined according to 
 \cite{Ito65,McKean75}
 \begin{equation}
 L(t)=D\lim_{\epsilon \rightarrow 0}\frac{1}{\epsilon}\int_0^t\Theta(\epsilon -X(\tau))d\tau.
 \end{equation}
 Second, the stochastic dynamics is described by a piecewise deterministic differential equation supplemented by three distinct jump processes, see equations (\ref{rPDMPa})--(\ref{rPDMPc}). The jumps represent tumbling, resetting and collision events, respectively.
 On the other hand, the stochastic dynamics of a Brownian particle with resetting on the half-line is described by a jump-diffusion SDE of the form
 \numparts
 \begin{eqnarray}
\label{BMa}
dX(t)&=\sqrt{2D}dW(t)+(x_0-X(t^-))d\overline{N}(t)+dL(t),\\
 \label{BMb}
dL(t)&= D\delta(X(t)) dt,
 \end{eqnarray}
 \endnumparts
 with $W(t)$ a Brownian motion, $D$ a diffusivity, and $\overline{N}(t)$ defined in equation (\ref{dNdefr}). Irrespective of the underlying stochastic process, one of the main steps in the encounter-based framework is deriving the evolution equation for the local time propagator. One principled way to achieve this is to use stochastic calculus and a version of It\^o's lemma as shown in section 3 for the RTP. (The analogous construction for Brownian motion is carried out in Ref. \cite{Bressloff24a}.)
 
 \begin{figure}[h!]
  \centering
   \includegraphics[width=12cm]{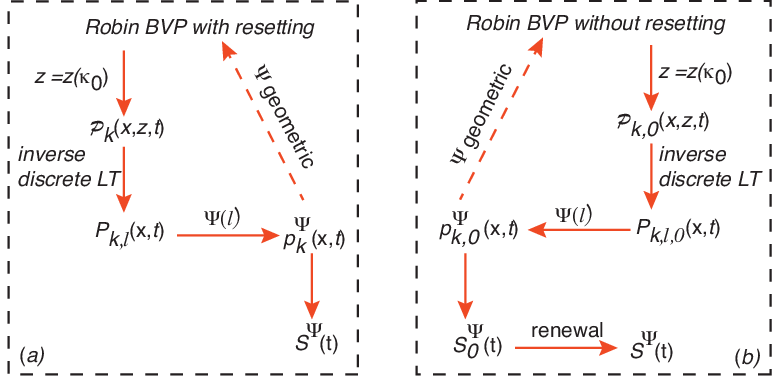}
  \caption{Summary diagrams of the encounter-based framework for an RTP on the half-line with a partially absorbing non-sticky wall and stochastic resetting. (a) Position and velocity resetting. Solve the Robin BVP with resetting and perform an inverse Laplace transform to determine the corresponding propagator $P_{k,l}$ where $k$ and $\ell$ label the current velocity state $\sigma(t)$ and the discrete local time $L(t)$. This then yields $p_k^{\Psi}=\int \Psi(\ell)P_{k,l}dl$ and the survival probability $S^{\Psi}$. (b) Position, velocity and local time resetting. First solve the Robin BVP without resetting (indicated by the subscript $0$) and perform an inverse Laplace transform to determine the corresponding propagator $P_{k,l,0}$. This then yields $p_{k,0}^{\Psi}=\int \Psi(\ell)P_{k,l,0}dl$ and the survival probability $S_0^{\Psi}$ without resetting. Finally, use a renewal equation to determine the survival probability with resetting.}
  \label{fig8}
  \end{figure}

The mathematical schemes used in our analysis of an RTP on the half-line with a partially absorbing boundary and stochastic resetting are summarized in Fig. \ref{fig8}. In the absence of local time resetting, we calculated the local time propagator with position and velocity resetting and used this to determine the survival probability $S^{\Psi}(t)$ for a given threshold distribution $\Psi$. (The survival probability does not satisfy a renewal equation relating it to the survival probability without resetting.) On the other hand, in the case where the local time also resets, we calculated the local time propagator and survival probability without resetting and used a renewal equation to determine $S^{\Psi}(t)$. We found that the MFPT in the first resetting scenario only depends on $\Psi$ via the mean local time threshold $\E[\ell]$, whereas in the second resetting scenario the MFPT depended on the full statistics of $\Psi$.

There are several possible future directions regarding encounter-based models of RTPs with resetting. First, we could consider a higher-dimensional model in which the RTP switches between velocity states in either a discrete number or continuum of different directions \cite{Mori20,Santra20,Santra20a}. This also raises the possibility of a spatially heterogeneous reactive surface. A second extension would be to allow the RTP to temporarily stick to the boundary. As shown in Ref. \cite{Bressloff22f} for an RTP without resetting, the contact time is given by the amount of time the particle is stuck to the surface, which can be equated with the occupation time of a boundary layer close to the surface.

\setcounter{equation}{0}
\renewcommand{\theequation}{A.\arabic{equation}}
\section*{Appendix A: Calculation of the survival probability without local time resetting}

In this appendix we solve the Laplace transformed Robin BVP given by equations (\ref{rLT1}) and (\ref{rLT2}) in order to determine the survival probability without local time resetting.

\subsection*{Case $\sigma_0=1$}  
First suppose that $\sigma_0=1$. Equation (\ref{rLT1}) is then equivalent to the pair of equations
\begin{eqnarray}
\label{CKLTa}
\fl & v\frac{\partial \calPP_{1}(x,z,s)}{\partial x}+(\alpha +s+r)\calPP_{1}(x,z,s)-\alpha \calPP_{-1}(x,z,s) \nonumber \\
\fl  &\qquad =\left (1+r\widetilde{\calS}(z,s)\right )\delta(x-x_0),\\
\fl & -v\frac{\partial \calPP_{-1}(x,z,s)}{\partial x}+(\alpha +s+r)\calPP_{-1}(x,z,s) -\alpha\calPP_{1}(x,z,s) =0.\label{CKLTb}
\end{eqnarray}
Introduce the Green's function $G_1(x,x_0;s)$ with
\begin{eqnarray}
\fl  \frac{\partial G_1(x,x_0;s)}{\partial x}+\lambda(s) G_1(x,x_0;s)=\delta(x-x_0),\quad G_1(0,x_0;s)=0
 \end{eqnarray}
 and
 \begin{equation}
\lambda(s)=\lambda_0(s+r)\equiv\frac{\alpha +s+r}{v}.
 \end{equation}
Solving for $G_1$ gives
\begin{equation}
G_1(x,x_0;s)=\e^{-\lambda(s)(x-x_0)}\Theta(x-x_0),
\end{equation}
with $\Theta(x)=1$ for $x>0$ and  zero otherwise. Substituting
\begin{equation}
\label{p1LT}
\calPP_{1}(x,z,s)=\frac{1}{v}\left (1+r\widetilde{\calS}(z,s)\right )G_1(x,x_0;s)+\calQQ_1(x,z,s)
\end{equation}
 into equation (\ref{CKLTa}) implies that
 \begin{equation}
 \frac{\partial \calQQ_1(x,z,s)}{\partial x}+\lambda(s) \calQQ_1(x,z,s)=\frac{\alpha}{v}\calPP_{-1}(x,z,s).
 \label{qLT}
 \end{equation}

The next step is to substitute the solution (\ref{p1LT}) into (\ref{CKLTb}):
\begin{eqnarray}
\label{PPm1}
 \fl -\frac{\partial \calPP_{-1}(x,z,s)}{\partial x}+\lambda(s) \calPP_{-1}(x,z,s)=\frac{\alpha}{v}\bigg \{\frac{1}{v}\left (1+r\widetilde{\calS}(z,s)\right )G_1(x,x_0;s)+\calQQ_1(x,z,s)\bigg \}.\nonumber \\ \fl
 \end{eqnarray}
 Applying the operator $\lambda(s)+\partial/\partial x$ to both sides and rearranging, we have
 \begin{eqnarray}
\fl  &\frac{\partial^2 \calPP_{-1}(x,z,s)}{\partial x^2}-\Lambda(s)^2\calPP_{-1}(x,z,s)
   =-\frac{\alpha}{v^2}\left (1+r\widetilde{\calS}(z,s)\right )\delta(x-x_0),
\end{eqnarray}
where
\begin{equation}
\Lambda(s)= \frac{1}{v}\sqrt{(s+r)(2\alpha+s+r)}\equiv \Lambda_0(s+r).
\end{equation}
We have used equation (\ref{Lam}). It follows that
\begin{equation}
\calPP_{-1}(x,z,s)=\frac{\alpha}{v^2}\left (1+r\widetilde{\calS}(z,s)\right )G(x,x_0;s)+\calQQ_{-1}(x,z,s),
\end{equation}
where
\begin{eqnarray}
 &\frac{\partial^2 \calQQ_{-1}(x,z,s)}{\partial x^2}-\Lambda(s+r)^2\calQQ_{-1}(x,z,s)
   =0,\ x >0, 
\end{eqnarray}
with $\calQQ_{1}(0,z,s)=z\calQQ_{-1}(0,z,s)$ and
\begin{equation}
\frac{\partial^2 G}{\partial x^2}-\Lambda^2(s)G=-\delta(x-x_0),\quad G(0,x_0;s)=0.
\end{equation}
The Green's function has the solution
\begin{equation}
 G(x,x_0;s)=\frac{1}{2\Lambda(s)}\left [\e^{-\Lambda(s) |x-x_0|}-\e^{-\Lambda(s) (x+x_0)}\right ],
 \label{G}
 \end{equation}
whereas
\begin{equation}
\calQQ_{-1}(x,z,s)=A(z,s)\e^{-\Lambda(s)x}.
\end{equation}
The unknown coefficient $A(z,s)=\calQQ_{-1}(0,z,s)=z^{-1}\calQQ_{1}(x,z,s)$ will be determined below.

We now substitute for $\calPP_{-1}$ in equation (\ref{qLT}) and solve using an integrating factor:
\begin{eqnarray*}
\fl \calQQ_1(x,z,s)&=\e^{-\lambda(s)x}\calQQ_1(0,z,s)+A(z,s) \frac{\alpha}{v}\int_{0}^x\e^{-\lambda(s)(x-y)}\e^{-\Lambda(s)y}dy \\
\fl &+\frac{1}{2\Lambda(s)}\frac{\alpha^2}{v^3}\left (1+r\widetilde{\calS}(z,s)\right )\int_{0}^x\e^{-\lambda(s)(x-y)}\left [\e^{-\Lambda(s) |y-x_0|}-\e^{-\Lambda(s) (y+x_0)}\right ]dy.\nonumber
\end{eqnarray*}
We evaluate each of the integrals as follows.
\begin{eqnarray*}
{\mathcal I}_1(x,s):=\int_{0}^x\e^{-\lambda(s) (x-y)}\e^{-\Lambda(s) y}dy=\frac{\e^{-\Lambda(s) x}-\e^{-\lambda(s) x}}{\lambda(s) -\Lambda(s) }.
\end{eqnarray*}
 If $x>x_0$ then
\begin{eqnarray*}
 \fl {\mathcal I}_>(x,s)&:=\int_{0}^x\e^{-\lambda(s)(x-y)}\e^{-\Lambda(s) |y-x_0|}
\nonumber \\
\fl  &=\e^{-\lambda(s) x}\e^{-\Lambda(s) x_0}\int_{0}^{x_0} \e^{(\lambda(s)+\Lambda(s))y}dy+\e^{-\lambda(s) x}\e^{\Lambda(s) x_0}\int_{x_0}^{x} \e^{(\lambda(s)-\Lambda(s))y}dy\nonumber\\
 \fl &=\frac{\e^{-\Lambda(s) (x-x_0)}}{\lambda(s)-\Lambda(s)}-\frac{\e^{-\lambda(s) x-\Lambda(s) x_0}}{\lambda(s)+\Lambda(s)}- \frac{2\Lambda(s)\e^{-\lambda(s) (x-x_0)}}{\lambda(s)^2-\Lambda(s)^2}.
\end{eqnarray*}
On the other hand, if $x <x_0$ then
\begin{eqnarray*}
\fl {\mathcal I}_<(x,s)&=\e^{-\lambda(s) x}\e^{-\Lambda(s) x_0}\int_{0}^{x} \e^{(\lambda(s)+\Lambda(s))y}dy=\frac{\e^{-\Lambda(s) |x-x_0|}}{\lambda(s)+\Lambda(s)}-\frac{\e^{-\lambda(s) x-\Lambda(s) x_0}}{\lambda(s)+\Lambda(s)}.
\end{eqnarray*}
Finally,
\begin{eqnarray*}
\fl {\mathcal I}_2(x,s)&=\int_{0}^x\e^{-\lambda(s) (x-y)}\e^{-\Lambda(s)  (y+x_0)}dy\nonumber \\
\fl &=\e^{-\lambda(s) x}\e^{-\Lambda(s) x_0}\int_{0}^{x} \e^{(\lambda(s)-\Lambda(s))y}dy=\frac{\e^{-\Lambda(s) (x+x_0)}}{\lambda(s)-\Lambda(s)}-\frac{\e^{-\lambda(s) x-\Lambda(s) x_0}}{\lambda(s)-\Lambda(s)}.
\end{eqnarray*}
Combining all of our results leads to the following solution in Laplace space:
\begin{eqnarray}
\fl \calQQ_1(x,z,s)&=\e^{-\lambda(s)x}\calQQ_1(0,z,s)+A(z,s) \frac{\alpha}{v}{\mathcal I}_1(x,s) +\frac{1}{2\Lambda(s)}\frac{\alpha^2}{v^3}\left (1+r\widetilde{\calS}(z,s)\right ) \\
\fl &\hspace{3cm}\times \bigg [ {\mathcal I}_>(x,s)\Theta(x-x_0)+ {\mathcal I}_<(x,s)\Theta(x_0-x)- {\mathcal I}_2(x,s))\bigg ].\nonumber
\end{eqnarray}

The final step is to determine the unknown coefficient $A(z,s)$ by plugging in our solutions into the first-order equation (\ref{PPm1}) for $x<x_0$:
\begin{eqnarray}
\fl &\frac{1}{2\Lambda}\frac{\alpha}{v^2}\left (1+r\widetilde{\calS}(z,s)\right )\left [[\lambda-\Lambda]\e^{-\Lambda(x_0-x)}-[\lambda+\Lambda]\e^{-\Lambda(x_0+x)}\right]+(\lambda+\Lambda) A(z,s)\e^{-\Lambda x}\nonumber\\
\fl &=\frac{\alpha}{v}A(z,s)\left [   z\e^{-\lambda x}+\frac{\alpha}{v}\frac{\e^{-\Lambda x}-\e^{-\lambda x}}{\lambda -\Lambda }
\right ]  \nonumber \\
\fl  &\quad+\frac{1}{2\Lambda}\frac{\alpha^3}{v^4}\left (1+r\widetilde{\calS}(z,s)\right )\bigg [ \frac{\e^{-\Lambda(x_0-x)}}{\lambda+\Lambda}-\frac{\e^{-\lambda x-\Lambda x_0}}{\lambda+\Lambda} - \frac{\e^{-\Lambda (x+x_0)}}{\lambda -\Lambda }+\frac{\e^{-\lambda  x-\Lambda  x_0}}{\lambda -\Lambda }\bigg ].
\end{eqnarray}
Using the identity $\lambda(s)^2-\Lambda(s)^2=\alpha^2/v^2$, we find that all terms in $\e^{\pm \Lambda x}$ cancel. Equating all the remaining terms involving the factor $\e^{-\lambda x}$ gives
\begin{eqnarray}
  \frac{\alpha}{v}A(z,s)\left [   z-\frac{\alpha}{v[\lambda -\Lambda] }
\right ]=-\frac{\alpha}{v^2}\left (1+r\widetilde{\calS}(z,s)\right )  \e^{-\Lambda x_0}.
\label{A}
\end{eqnarray}
From equation (\ref{tilS}) we note that
\begin{eqnarray}
\label{AA}
s \widetilde{\calS}(z,s)=1-(1-z)vA(z,s).
\end{eqnarray}
Using the fact that the particle is eventually absorbed when $r>0$ and $z<1$, we have
\begin{equation}
\lim_{t\rightarrow\infty} \calS(z,t)=\lim_{s\rightarrow 0} s \widetilde{\calS}(z,s)=0.
\end{equation}
Hence,
\begin{equation}
\lim_{s\rightarrow 0} A(z,s)=\frac{1}{v[1-z]}.
\end{equation}
Taking the limit $s\rightarrow 0$ of equation (\ref{A}) then yields equation (\ref{A0}).

\subsection{Case $\sigma_0=-1$}  
We now follow similar steps to solve the propagator equation CK equation (\ref{rLT1}) with $\sigma_0=-1$: 
\begin{eqnarray}
\label{mCKLTa}
 \fl & v\frac{\partial \calPP_{1}(x,z,s)}{\partial x}+(\alpha +s+r)\calPP_{1}(x,z,s)-\alpha \calPP_{-1}(x,z,s) =0,\\
\fl & -v\frac{\partial \calPP_{-1}(x,z,s)}{\partial x}+(\alpha +s+r)\calPP_{-1}(x,z,s) -\alpha\calPP_{1}(x,z,s) \nonumber \\
\fl &\quad =\left (1+r\widetilde{\calS}(z,s)\right )\delta(x-x_0).\label{mCKLTb}
\end{eqnarray}
Introduce the Green's function $G_{-1}(x,x_0;s)$ with
\begin{eqnarray}
\fl  \frac{\partial G_{-1}(x,x_0;s)}{\partial x}-\lambda(s) G_{-1}(x,x_0;s)=-\delta(x-x_0). 
 \end{eqnarray}
Solving for $G_1$ gives
\begin{equation}
G_{-1}(x,x_0;s)=\e^{-\lambda(s)(x_0-x)}\Theta(x_0-x),\quad x\geq 0.
\end{equation}
Substituting
\begin{equation}
\label{mp1LT}
\calPP_{-1}(x,z,s)=\frac{1}{v}\left (1+r\widetilde{\calS}(z,s)\right )G_{-1}(x,x_0;s)+\calQQ_{-1}(x,z,s)
\end{equation}
 into equation (\ref{mCKLTb}) implies that
 \begin{equation}
 \frac{\partial \calQQ_{-1}(x,z,s)}{\partial x}-\lambda(s) \calQQ_{-1}(x,z,s)=-\frac{\alpha}{v}\calPP_{1}(x,z,s).
 \label{mqLT}
 \end{equation}
Moreover, substituting the solution (\ref{mp1LT}) into (\ref{mCKLTa}):
\begin{eqnarray}
\label{mPPm1}
 &\frac{\partial \calPP_{1}(x,z,s)}{\partial x}+\lambda(s) \calPP_{1}(x,z,s)\\
  &=\frac{\alpha}{v}\bigg \{\frac{1}{v}\left (1+r\widetilde{\calS}(z,s)\right )G_{-1}(x,x_0;s)+\calQQ_{-1}(x,z,s)\bigg \}.\nonumber 
 \end{eqnarray}
 Applying the operator $-\lambda(s)+\partial/\partial x$ to both sides and rearranging, we have
 \begin{eqnarray}
\fl  &\frac{\partial^2 \calPP_{1}(x,z,s)}{\partial x^2}-\Lambda(s)^2\calPP_{1}(x,z,s)
   =-\frac{\alpha}{v^2}\left (1+r\widetilde{\calS}(z,s)\right )\delta(x-x_0),
\end{eqnarray}
and thus
\begin{equation}
\calPP_{1}(x,z,s)=\frac{\alpha}{v^2}\left (1+r\widetilde{\calS}(z,s)\right )G(x,x_0;s)+\calQQ_{1}(x,z,s),
\end{equation}
with $G$ given by equation (\ref{G}) and
\begin{equation}
\calQQ_{1}(x,z,s)=B(z,s)\e^{-\Lambda(s)x}.
\end{equation}
The unknown coefficient $B(z,s)$ satisfies the boundary condition (\ref{rLT2}) at $x=0$, that is, 
\begin{equation}
\fl B(z,s)=z\calPP_{-1}(0,z,s) =z\left [\calQQ_{-1}(0,z,s) +\frac{1}{v}\left (1+r\widetilde{\calS}(z,s)\right )\e^{-\lambda(s)x_0}\right ].
 \label{mrLT2}
\end{equation}
We now substitute for $\calPP_{1}$ in equation (\ref{mqLT}) and solve using an integrating factor:
\begin{eqnarray}
\fl &\calQQ_{-1}(x,z,s)\nonumber \\
\fl &=\e^{\lambda(s)x}\calQQ_{-1}(0,z,s)-B(z,s) \frac{\alpha}{v}\int_{0}^x\e^{\lambda(s)(x-y)}\e^{-\Lambda(s)y}dy\\
\fl &-\frac{1}{2\Lambda(s)}\frac{\alpha^2}{v^3}\left (1+r\widetilde{\calS}(z,s)\right )\int_{0}^x\e^{\lambda(s)(x-y)}\left [\e^{-\Lambda(s) |y-x_0|}-\e^{-\Lambda(s) (y+x_0)}\right ]dy.\nonumber 
\end{eqnarray}
The integral terms can be evaluated along identical lines to the previous case under the transformation ${\mathcal I}\rightarrow \overline{\mathcal I}$ with $\overline{\mathcal I}(\lambda)={\mathcal I}(-\lambda)$> hence,
\begin{eqnarray}
\label{ans}
\fl &\calQQ_{-1}(x,z,s)\nonumber \\
\fl &=\e^{\lambda(s)x}\calQQ_{-1}(0,z,s)-B(z,s) \frac{\alpha}{v}\overline{\mathcal I}_1(x,s)  \\
\fl &-\frac{1}{2\Lambda(s)}\frac{\alpha^2}{v^3}\left (1+r\widetilde{\calS}(z,s)\right )\bigg [ \overline{\mathcal I}_>(x,s)\Theta(x-x_0)+\overline {\mathcal I}_<(x,s)\Theta(x_0-x)- \overline{\mathcal I}_2(x,s))\bigg ].\nonumber
\end{eqnarray}


The final step is to determine the unknown coefficient $B(z,s)$ by plugging in our solutions into the first-order equation (\ref{mPPm1}) for $x<x_0$:
\begin{eqnarray}
\label{ans2}
\fl &\frac{1}{2\Lambda}\frac{\alpha}{v^2}\left (1+r\widetilde{\calS}(z,s)\right )\left [[\lambda+\Lambda]\e^{-\Lambda(x_0-x)}-[\lambda-\Lambda]\e^{-\Lambda(x_0+x)}\right]+(\lambda-\Lambda) B(z,s)\e^{-\Lambda x}\nonumber\\
\fl &=\frac{\alpha}{v}B(z,s)\left [   z^{-1}\e^{\lambda x}+\frac{\alpha}{v}\frac{\e^{-\Lambda x}-\e^{\lambda x}}{\lambda +\Lambda }
\right ] 
\nonumber \\
\fl  &\quad+\frac{1}{2\Lambda}\frac{\alpha^3}{v^4}\left (1+r\widetilde{\calS}(z,s)\right )\bigg [ \frac{\e^{-\Lambda(x_0-x)}}{\lambda-\Lambda}-\frac{\e^{\lambda x-\Lambda x_0}}{\lambda-\Lambda} - \frac{\e^{-\Lambda (x+x_0)}}{\lambda +\Lambda }+\frac{\e^{\lambda  x-\Lambda  x_0}}{\lambda +\Lambda }\bigg ].
\end{eqnarray}
Using the identity $\lambda(s)^2-\Lambda(s)^2=\alpha^2/v^2$, we again find that all terms in $\e^{\pm \Lambda x}$ cancel. Equating all the remaining terms involving the factor $\e^{\lambda x}$ gives
\begin{eqnarray}
 \frac{\alpha}{v}B(z,s)\left [   z^{-1}-\frac{\alpha}{v[\lambda +\Lambda] }
\right ]=\frac{\alpha}{v^2}\left (1+r\widetilde{\calS}(z,s)\right )  \e^{-\Lambda x_0}.
\label{B}
\end{eqnarray}
From equation (\ref{tilS}) and the identity $\calPP_{-1}(0,z,s)=z^{-1}B(z,s)$, we have (for initial data $(x_0,-1)$)
\begin{eqnarray}
\label{BB}
s \widetilde{\calS}(z,s)=1-\frac{1-z}{z}vB(z,s).
\end{eqnarray}
Again using the fact that the particle is eventually absorbed when $r>0$ and $z<1$, we have
\begin{equation}
\lim_{t\rightarrow\infty} \calS(z,t)=\lim_{s\rightarrow 0} s \widetilde{\calS}(z,s)=0.
\end{equation}
Hence,
\begin{equation}
\lim_{s\rightarrow 0} B(z,s)=\frac{z}{v[1-z]}.
\end{equation}
Taking the limit $s\rightarrow 0$ of equation (\ref{B}) then yields equation (\ref{B0}).

\end{document}